\documentclass[%
 aip,
 amsmath,amssymb,
reprint,%
]{revtex4-1}

\usepackage{graphicx}
\usepackage{dcolumn}
\usepackage{bm}
\usepackage{float}
\usepackage[utf8]{inputenc}
\usepackage[T1]{fontenc}
\usepackage{mathptmx}
\usepackage{tikz}
\usetikzlibrary{arrows.meta,calc}
\usepackage{booktabs} 
\usepackage{braket}
\usepackage{multirow}
\usepackage{hyperref}

\usepackage{placeins}
\begin{document}

\preprint{AIP/123-QED}

\title{Convergence to the fixed-node limit in deep variational Monte Carlo}

\author{Z. Schätzle}
\affiliation{FU Berlin, Department of Mathematics and Computer Science,
Arnimallee 6, 14195 Berlin, Germany}
\author{J. Hermann}
\affiliation{FU Berlin, Department of Mathematics and Computer Science,
Arnimallee 6, 14195 Berlin, Germany}
\affiliation{TU Berlin, Machine Learning Group, Marchstr. 23, 10587 Berlin, Germany}
\author{F. Noé}
\email{frank.noe@fu-berlin.de}
\affiliation{FU Berlin, Department of Mathematics and Computer Science,
Arnimallee 6, 14195 Berlin, Germany}
\affiliation{FU Berlin, Department of Physics, Arnimallee 14, 14195 Berlin, Germany}
\affiliation{Rice University, Department of Chemistry, Houston, TX 77005, USA}

\begin{abstract}
Variational quantum Monte Carlo (QMC) is an ab-initio method for solving the electronic Schrödinger equation that is exact in principle, but limited by the flexibility of the available ansatzes in practice.
The recently introduced deep QMC approach, specifically two deep-neural-network ansatzes PauliNet and FermiNet, allows variational QMC to reach the accuracy of diffusion QMC, but little is understood about the convergence behavior of such ansatzes.
Here, we analyze how deep variational QMC approaches the fixed-node limit with increasing network size.
First, we demonstrate that a deep neural network can overcome the limitations of a small basis set and reach the mean-field complete-basis-set limit.
Moving to electron correlation, we then perform an extensive hyperparameter scan of a deep Jastrow factor for LiH and H$_4$ and find that variational energies at the fixed-node limit can be obtained with a sufficiently large network.
Finally, we benchmark mean-field and many-body ansatzes on H$_2$O, increasing the fraction of recovered fixed-node correlation energy of single-determinant Slater--Jastrow-type ansatzes by half an order of magnitude compared to previous variational QMC results and demonstrate that a single-determinant Slater--Jastrow--backflow version of the ansatz overcomes the fixed-node limitations.
This analysis helps understanding the superb accuracy of deep variational ansatzes in comparison to the traditional trial wavefunctions at the respective level of theory, and will guide future improvements of the neural network architectures in deep QMC\@.
\end{abstract}
\maketitle

\section{\label{sec:Introduction}Introduction}
The fundamental problem in quantum chemistry is to solve the electronic Schrödinger equation as accurately as possible at a manageable cost.
Variational quantum Monte Carlo (variational QMC, or VMC for short) is an ab-initio method based on the stochastic evaluation of quantum expectation values that scales favorably with system size and provides explicit access to the wavefunction~\cite{foulkesQuantumMonteCarlo2001a}.
Although exact in principle, VMC strongly depends on the quality of the trial wavefunction, which determines both efficiency and accuracy of the computation and typically constitutes the limiting factor of VMC calculations.

Recently, deep QMC has been introduced.
Deep QMC involves a new class of ansatzes that complement traditional trial wavefunctions with the expressiveness of deep neural networks (DNNs).
This ab-initio approach is orthogonal to the supervised learning of electronic structure that requires external datasets~\cite{SchuttNC19,GasteggerJCP20}.
The use of neural network trial wavefunctions has been pioneered for spin lattice systems~\cite{carleoSolvingQuantumManybody2017} and later generalized to molecules in second quantization~\cite{chooFermionicNeuralnetworkStates2020a}.
The first application to molecules in real space was a proof-of-principle effort, but did not reach the accuracy close to traditional VMC~\cite{hanSolvingManyelectronSchrodinger2019a}.
The DNN architectures PauliNet and FermiNet advanced the real-space deep QMC approach~\cite{hermannDeepneuralnetworkSolutionElectronic2020,pfauInitioSolutionManyelectron2020}, increasing the accuracy to state-of-the-art levels and beyond.
Demonstrating very high accuracy with far fewer determinants than traditional counterparts, these deep neural network trial wavefunctions provide an alternative to increasing the number of Slater determinants, thus potentially improving the unfavorable scaling with respect to the number of electrons that complicates accurate calculations for large systems.
Application of the deep QMC method to many-particle quantum systems other than electrons is also possible~\cite{Adams20}.

Currently, there is little understanding of why these DNN wavefunctions work well and how their individual components contribute to the approximation of the ground-state wavefunction and energy. Examining their expressive power and measuring their accuracy in comparison to traditional approaches is essential to establish neural-network trial wavefunctions as a standard technique in VMC and to guide further development.

\begin{figure}[t]
    \centering
    \vspace{5mm}
    \includegraphics{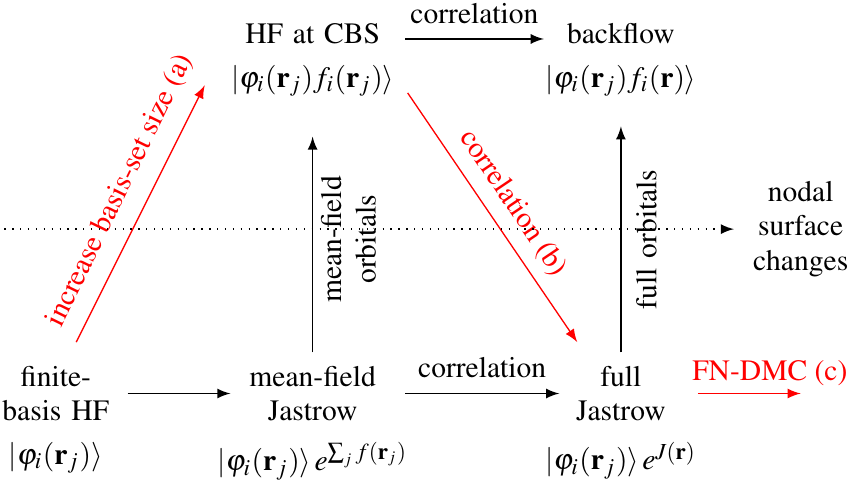}
\caption{\textbf{Hierarchy of single-determinant ansatzes in QMC.} 
The starting point of a finite-basis Hartree–Fock (HF) calculation can be extended by a “mean-field” Jastrow factor to improve the one-electron density of the ansatz. From that point, the ansatz can be improved in one of two directions---by modifying the orbitals (bottom--top) or introducing electron correlation (left--right). The red pathway
shows a standard approach in traditional QMC and is the path we pursue in our analysis.
\label{fig:sketch}}
\end{figure}

Here, we identify a hierarchy of model ansatzes based on the traditional VMC methodology (Fig.~\ref{fig:sketch}) that enables us to distinguish the effects of improving single-particle orbitals and adding correlation in the symmetric part of the wavefunction ansatz. This is of particular interest in the context of discriminating these improvements from reducing the energy by solving the intricate problem of missing many-body effects in the nodal surface. \\

The trial wavefunctions in QMC are typically constructed by combining a symmetric Jastrow factor with an antisymmetric part that implements the Pauli exclusion principle for fermions by specifying the nodal surface of the ansatz---the hyperplane in the space of electron coordinates, $\mathbf r~=~(\mathbf r_1,\ldots,\mathbf r_N)$, on which the wavefunction changes sign.
Expressing the antisymmetric part as a linear combination of Slater determinants gives rise to the ansatz of the Slater--Jastrow--backflow-type that comprises most VMC ansatzes, including the deep variants PauliNet and FermiNet,
\begin{equation}
\psi(\mathbf r)
  = \underset{\mathrm{symmetric}}{\underbrace{\mathrm e^{J(\mathbf r)}}}
  \underset{\mathrm{antisymmetric}}{\underbrace{
  \textstyle{\sum}_p c_p
  \det[\mathbf A_p^\uparrow(\mathbf r)]
  \det[\mathbf A_p^\downarrow(\mathbf r)]}}
\label{eq:general_ansatz}
\end{equation}
The ability of neural networks to represent antisymmetric (wave) functions has been also explored theoretically~\cite{Han19,Hutter20}.

Traditionally, Slater determinants are antisymmetrized product states constructed from single-particle molecular orbitals, which are expressed in a one-electron basis set consisting of basis functions $\phi_{k}$,
\begin{equation}
    A_{ij}=\varphi_i(\mathbf r_j) = \sum\nolimits_{k} c_{ik} \phi_{k}(\mathbf r_j) 
\end{equation}
Employing such basis sets transforms the problem of searching over infinitely many functions into a problem of searching over coefficients in a system of equations, which can be solved by means of linear algebra applying for instance the Hartree--Fock (HF), the multi-configurational self-consistent field (MCSCF), or the full configuration interaction (FCI) method. The projection comes at the cost of introducing the finite-basis-set error, which completely vanishes only in the limit of infinitely many basis function---the complete-basis-set (CBS) limit  (Fig.~\ref{fig:sketch}a). Finite-basis-set errors are inherent to the second-quantized representation, which nevertheless provides an alternative platform to introduce deep learning to quantum chemistry~\cite{chooFermionicNeuralnetworkStates2020a}.

The real-space formulation of VMC allows to introduce explicit electron correlation efficiently by modelling many-body interactions with a Jastrow factor (Fig.~\ref{fig:sketch}b). The Jastrow factor is a symmetric function of the electron coordinates that traditionally involves an expansion in one, two, and three-body terms~\cite{drummondJastrowCorrelationFactor2004a}. Although strongly improving the ansatz, traditional Jastrow factors do not have sufficient expressiveness to reach high accuracy and an initial VMC calculation is typically followed by a computationally demanding fixed-node diffusion QMC (FN-DMC) simulation (Fig.~\ref{fig:sketch}c), which eventually projects out the exact solution for the given nodal surface---the fixed-node limit~\cite{needsContinuumVariationalDiffusion2009}. 
DMC is based on the imaginary-time Schrödinger equation and offers yet another entry point for the use of neural networks to represent quantum states~\cite{BarrPMLR20,Han20}.

The nodal surface of the trial wavefunctions can be improved by increasing the number of determinants or by applying the backflow technique, transforming single-particle orbitals to many-body orbitals under consideration of the symmetry constraints. These are key concepts to efficiently reach very high accuracy with VMC and integral features of deep QMC. Using multiple determinants, applying the backflow technique and modifying the symmetric component of the ansatz at the same time however makes it difficult identify the contributions of each individual part. Benchmarking deep QMC ansatzes in conceptually simpler contexts confirms their correct functionality and helps to achieve a better understanding.

In this paper we take a closer look at how neural networks compensate for errors arising from finite basis sets and demonstrate convergence to the fixed-node limit within the VMC framework by systematically increasing the expressiveness of a deep Jastrow factor. For the sake of disentangling the individual contributions to the overall accuracy, we conduct our analysis mainly with Slater--Jastrow-type trial wavefunctions with an antisymmetric part consisting of a single determinant, that is, with ansatzes possessing a mean-field nodal surface. We compare neural-network variants with traditional functional forms, as well as with DMC results. In particular we investigate the PauliNet, a recently proposed neural-network trial wavefunction\cite{hermannDeepneuralnetworkSolutionElectronic2020}. PauliNet combines ideas from conventional trial wavefunctions, such as a symmetric Jastrow factor, a generalized backflow transformation, multi-determinant expansions, quantum chemistry baselines, and an explicit implementation of physical constraints of ground-state wavefunctions. Since PauliNet is a powerful instance of the general ansatz in (\ref{eq:general_ansatz}), we can obtain traditional types of QMC ansatzes at different levels of the theory by deactivating certain trainable parts of PauliNet. The hierarchy of ansatzes sketched in Fig.~\ref{fig:sketch} maps restricted single-determinant versions of PauliNet and their eventual expressiveness in the context of the traditional single-determinant VMC approach. The incentive of implementing restricted variants of PauliNet is to test the behavior of the ansatz in settings that are well solved by existing methods and investigate the expressiveness of the individual components of PauliNet on well-defined subproblems. These restricted variants however are not intended to be used in order to achieve best accuracy, which is attained when taking advantage of the full flexibility of the PauliNet ansatz as demonstrated previously \cite{hermannDeepneuralnetworkSolutionElectronic2020}.

The rest of the paper is organized as follows. In Sec.~\ref{sec:Theory} we review the general PauliNet ansatz and show how different levels of the model hierarchy (Fig.\ref{fig:sketch}) can be obtained. In Sec.~\ref{sec:Results} we use these instances of PauliNet to investigate several subproblems of the fixed-node limit within the deep QMC approach. First we demonstrate that DNNs can be employed to correct the single-particle orbitals of a HF calculation in a small basis and obtain energies close to the CBS limit. Next we benchmark the deep Jastrow factor. We start by applying it to two node-less test systems, H$_2$ and He, where results within five significant digits of the exact energy are achieved. Next we conduct an extensive hyperparameter search for two systems with four electrons, LiH and the H$_4$ rectangle, revealing that the expressiveness of the ansatz can be systematically increased to converge to the fixed-node limit imposed by the employed antisymmetric ansatz. We further explore the convergence aspect by sampling the dipole moment for the LiH ansatzes and evaluating energy differences for two configurations of the hydrogen rectangle. Thereafter we show the size consistency of the method, examining the optimization of the deep Jastrow factor for systems of non-interacting molecules (H$_2$-H$_2$ and LiH-H$_2$). Finally we test various single-determinant variants of PauliNet in an analysis of the water molecule and compare them to traditional trial wavefunctions. Section~IV discusses the results.

\section{\label{sec:Theory}Theory and methods}
\subsection{\label{subsec:PauliNet}PauliNet}
The central object of our investigation is PauliNet, a neural-network trial wavefunction of the form in \eqref{eq:general_ansatz}. PauliNet extends the traditional multi-determinant Slater--Jastrow--backflow-type trial wavefunctions\cite{lopezriosInhomogeneousBackflowTransformations2006a}, retaining physically motivated structural features while replacing ad-hoc parameterizations with highly expressive DNNs,
\begin{gather}
\label{eq:ansatz}
\psi_{\boldsymbol\theta}(\mathbf r)
  =\mathrm e^{\gamma(\mathbf r)+J_{\boldsymbol\theta}(\mathbf r)}
  \sum_p c_p
  \det[\tilde\varphi_{\boldsymbol\theta,{\mu_p}i}^\uparrow(\mathbf r)]
  \det[\tilde\varphi_{\boldsymbol\theta,{\mu_p}i}^\downarrow(\mathbf r)]
  \\
\label{eq:backflow}
\tilde\varphi_{\boldsymbol\theta,\mu i}(\mathbf r)
  =\varphi_\mu(\mathbf r_i)f^{\boldsymbol \otimes}_{\boldsymbol\theta,\mu i}(\mathbf r)+f^{\boldsymbol \oplus}_{\boldsymbol\theta,\mu i}(\mathbf r)
\end{gather}
The ansatz consists of a linear combination of Slater determinants of molecular single-particle orbitals $\varphi_\mu$ corrected by a generalized backflow transformation $\mathbf{f}_{\boldsymbol\theta}$ and of a Jastrow factor $J_{\boldsymbol\theta}$. The DNN components are indicated by the $\boldsymbol\theta$ subscript, denoting the trainable parameters of the involved neural networks. The expansion coefficients $c_p$ and the single-particle orbitals are initialized from a preceding standard quantum-chemistry calculation (HF or MCSCF). The analytically known electron--nucleus and electron--electron cusp conditions \cite{katoEigenfunctionsManyparticleSystems1957a} are enforced within the orbitals $\varphi_\mu$ and as a fixed part $\gamma$ of the Jastrow factor, respectively. The correct cusps are maintained by designing the remaining trial wavefunction architecture to be cusp-less.

Both backflow transformation and Jastrow factor can introduce many-body correlation and are constructed in such a way that they preserve the anti-symmetry of the trial wavefunction. The Jastrow factor consists of a symmetric function, that is it retains the antisymmetry upon being invariant under the exchange of same-spin electrons. This however has the consequence that it scales the wavefunction without altering the nodes of the ansatz. The backflow transformation on the other hand alters the nodal surface by acting on the orbitals directly. Traditionally the backflow correction introduces many-body correlation by assigning quasi-particle coordinates that get streamed through the original orbitals. PauliNet generalizes this concept, based on the observation that equivariance with respect to the exchange of electrons is a sufficient criterion to retain the antisymmetry of the Slater determinant, and considers the backflow correction as a many-body transformation of the orbitals themselves. In fact it has been shown that in principle a single Slater determinant with generalized orbitals is capable of representing any antisymmetric function, if the many-body orbitals are sufficiently expressive\cite{Hutter20}. Both Jastrow factor $J_{\boldsymbol\theta}$ and backflow transformation $\mathbf{f}_{\boldsymbol\theta}$ are obtained from a joint latent-space representation encoded by a graph-convolutional neural network. The network acts on the rotation- and translation-invariant representation of the system given by the fully-connected graph of distances between all electrons and nuclei. The latent-space many-body representation is designed to be equivariant under the exchange of same-spin electrons, which is used to construct the permutation-equivariant backflow transformation and the permutation-invariant Jastrow factor. Details on the graph-convolutional neural network architecture can be found in the \hyperref[appendix]{Appendix}. 
Combining an expansion in Slater-determinants with Jastrow factor and backflow transformation introduces multiple different ways to model many-body effects, helping to efficiently encode correlation in the ansatz by representing i.e. dynamic correlation explicitly while implementing static correlation with multiple determinants.

The PauliNet ansatz is optimized according to the standard VMC scheme \cite{foulkesQuantumMonteCarlo2001a} of minimizing its energy expectation value. This is based on the variational principle of quantum mechanics that guarantees the energy expectation value of any trial wavefunction to be lower-bounded by the ground-state energy, as long as the fermionic antisymmetry constraint is implemented,
\begin{equation}
    E_0=\min_\psi\langle \psi|\hat H|\psi\rangle \le\min_{\boldsymbol\theta}\langle \psi_{\boldsymbol\theta}|\hat H|\psi_{\boldsymbol\theta}\rangle,\quad\psi\in\mathcal{H}^-
\end{equation}
In VMC this expectation value is approximated by Monte Carlo integration,
\begin{equation}
\langle \psi_{\boldsymbol\theta}|\hat H|\psi_{\boldsymbol\theta}\rangle \approx\frac{1}{M}\sum_{k=1}^{M}\frac{\hat H\psi_\theta(\mathbf r_k)}{\psi_{\boldsymbol\theta}( \mathbf r_k)}, \quad \mathbf r_k \in \mathbb{R}^{3N} \sim|\psi_{\boldsymbol\theta}|^2
\end{equation}
In practice this gives rise to an alternating scheme of sampling electronic configurations according to the probability density associated with the trial wavefunction with a standard Langevin sampling approach and optimizing the parameters of this wavefunction by following their (stochastic) gradient with respect to estimates of the expectation value over small batches. For further details of the training methodology see ref.~\onlinecite{hermannDeepneuralnetworkSolutionElectronic2020}. 
Numerical calculations were carried out with the DeepQMC Python package \cite{janhermannDeepqmcDeepqmcDeepQMC2021}, with training hyperparameters as reported in Table~\ref{tab:hyperparameters_training}.

Next we show how to obtain the ansatzes of Fig.~\ref{fig:sketch} from the general PauliNet architecture and introduce the respective optimization problems to be solved. 
\subsection{Deep orbital correction}
\label{subsec:deep_orbital}
The simplest way to approach the quantum many-body problem is by considering a mean-field theory. The HF method gives the optimal mean-field solution within the space of the employed basis set. A mean-field variant of the PauliNet architecture can be used to account for finite-basis-set errors in the HF baseline, by introducing a real-space correction to the single-particle orbitals,
\begin{gather}
\label{eq:anastz_orbital_correction}
\psi_{\boldsymbol\theta}(\mathbf r)
  =\mathrm 
  \det[\tilde\varphi_{\boldsymbol\theta,\mu}^\uparrow(\mathbf r_i)]
  \det[\tilde\varphi_{\boldsymbol\theta,\mu}^\downarrow(\mathbf r_i)]
  \\
\label{eq:ansatz_mean-field_backflow}
\tilde\varphi_{\boldsymbol\theta,\mu}(\mathbf r_i)
  =\varphi_\mu(\mathbf r_i)f^{\boldsymbol \otimes}_{\boldsymbol\theta,\mu}\big(\mathbf x_i^{(L)}(\mathbf r_i)\big)+f^{\boldsymbol \oplus}_{\boldsymbol\theta,\mu}\big(\mathbf x_i^{(L)}(\mathbf r_i)\big)
\end{gather}
The functions $f^{\boldsymbol \otimes}_{\boldsymbol\theta}$ and $f^{\boldsymbol \oplus}_{\boldsymbol\theta}$ are implemented by DNNs that generate a multiplicative and an additive correction to the HF orbitals $\varphi_\mu$, respectively. Combining a multiplicative and an additive correction serves the practical purpose of facilitating the learning process, as the multiplicative correction has a strong effect where the value of the orbital is large, while the additive correction can alter the nodes of the molecular orbital. (In principle, an additive correction only would be a sufficient parameterization.) This approach is a special case of the generalized backflow transformation in \eqref{eq:backflow}, in which the backflow correction depends on the position of the $i$-th electron only. The single-particle representation $\mathbf x_i^{(L)}(\mathbf r_i)$ can be obtained by a slight modification of the graph-convolutional architecture as described in the \hyperref[appendix]{Appendix}. If Gaussian-type orbitals are used, it is common to correct the missing nuclear cusp at the coalescence points within the orbitals. We employ the cusp correction of \citet{maSchemeAddingElectron2005a} and construct the DNNs to be cusp-less. 
Though the DNN could in principle approximate the orbitals from scratch, providing the HF baseline that ensures the correct asymptotics and offers a good initial guess reduces the training cost and makes the training process more robust.
In the mean-field theory the HF energy at the CBS limit constitutes a benchmark for the best possible solution to the optimization problem.
\subsection{Deep Jastrow factor}
\label{subsec:deep_jastrow}
The Slater--Jastrow-type ansatz goes beyond the mean-field theory by introducing explicit electronic correlation. The symmetric Jastrow factor however cannot alter the nodal surface and the single-determinant Slater--Jastrow-type ansatz is therefore a many-body ansatz possessing a mean-field nodal surface,
\begin{equation}
\psi_{\boldsymbol\theta}(\mathbf r)
  =\mathrm e^{\gamma(\mathbf r)+J_{\boldsymbol\theta}(\mathbf r)}
  \det[\varphi_{{\mu}}^\uparrow(\mathbf{r}_i)]
  \det[\varphi_{{\mu}}^\downarrow(\mathbf{r}_i)]
\label{eq:ansatz_Jastrow}
\end{equation}
The deep Jastrow factor $J_{\boldsymbol\theta}$ is obtained from the latent-space many-body representation encoded by the graph-convolutional neural network described in the \hyperref[appendix]{Appendix},
\begin{equation}\label{eq:deep_jastrow}
  J_{\boldsymbol\theta}(\mathbf r):=\eta_{\boldsymbol\theta} \big(\textstyle\sum_i\mathbf x_i^{(L)}(\mathbf r)\big)
\end{equation}
To enforce the symmetry of the Jastrow factor, the permutation-equivariant many-body embeddings $\mathbf x_i^{(L)}$ are summed over the electrons to give permutation invariant features. These features serve as input to a fully-connected neural network $\eta_\theta$, which returns the final Jastrow factor. The process of obtaining the latent-space representation involves multiple smaller components, such as trainable arrays and fully-connected neural networks, whose full specification gives rise to a collection of hyperparameters that influence the expressiveness of the ansatz. A list of the components and the respective hyperparameters can be found in Table~\ref{tab:hyperparameters}.

Benchmarking Jastrow factors comes with the difficulty of distinguishing errors arising from the nodal surface from those present due to a lack of expressiveness in the Jastrow factor. The optimal energy of a Slater--Jastrow-type trial wavefunction however can be obtained with the FN-DMC algorithm, that gives the exact ground state of the Schrödinger equation under the fixed-node constraint of the antisymmetric part of the ansatz.

\subsection{Mean-field Jastrow factor}
\label{subsec:mean-field_jastrow}
We furthermore implement a mean-field Jastrow factor, which constitutes another point in the space of ansatz classes (Fig.~\ref{fig:sketch}),
\begin{equation}\label{eq:ansatz_mean-field_Jastrow}
  J_{\boldsymbol\theta}(\mathbf r):=\textstyle\sum_i\eta_{\boldsymbol\theta} \big(\mathbf x_i^{(L)}(\mathbf r_i)\big)
\end{equation}
The mean-field Jastrow factor can optimize the one-electron density of the ansatz without modifying the nodal surface or introducing correlation, making its variations a strict subset of the orbital correction. This equips us with an intermediate step in approaching the finite-basis-set limit, that can be used to relate the finite-basis-set error to the fixed-node error of the HF baseline. If the many-body Jastrow factor is used, the mean-field version is not needed, as it is implicitly included in the many-body version.
\section{\label{sec:Results}Results}

\subsection{Large basis sets are not necessary in DNN ansatzes}

\begin{figure}[t]
    \centering
    \includegraphics{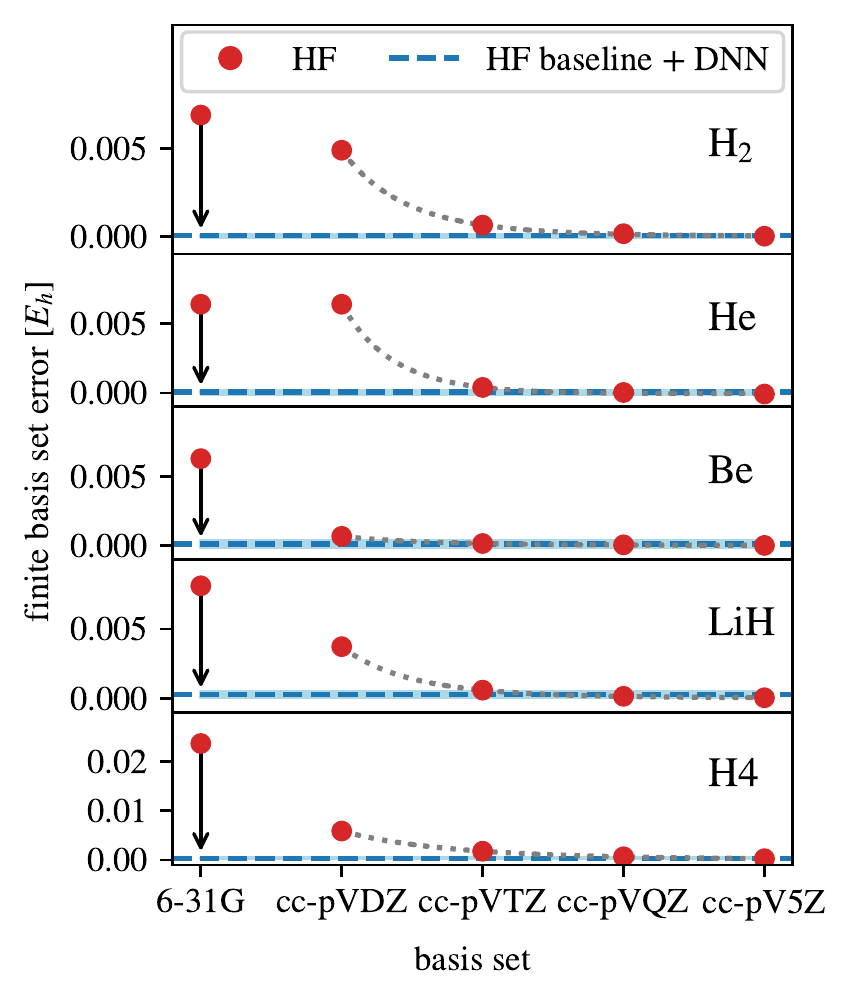}
    \caption{\label{fig:remove_bse} \textbf{Removing the basis-set error of a HF calculation}. The error with respect to the estimated CBS limit of HF calculations with increasingly large basis sets, as well as the result of employing a DNN to correct the single-particle orbitals of a HF calculation in a small basis (6-31G) are shown. The statistical error of the Monte Carlo integration is shown in light blue. The DNN is capable to correct deficiencies arising from the finite basis, producing energies close to the CBS limit.}
\end{figure}

We start from a HF baseline obtained in the small 6-31G basis set. Instead of introducing more basis functions, the PauliNet ansatz follows the alternative approach of correcting the orbitals directly in real space. We trained the mean-field variant of the PauliNet ansatz (Sec.~\ref{subsec:deep_orbital}) on H$_2$, He, Be, LiH and the hydrogen square H$_4$. For all five test systems we obtained energies close to the extrapolated CBS limit and recovered at least 97\% of the finite-basis-set error (Fig.~\ref{fig:remove_bse}). This shows that the use of a very small basis set for the baseline of PauliNet does not introduce any fundamental limitation to accuracy, because the neural network is able to correct it. We note that such an approach to the CBS limit is practical only within the context of the full PauliNet, not as a standalone technique to replace large basis sets in quantum chemistry.
\subsection{Exact solutions for two electron systems}
\begin{figure}[t]
    \centering
    \includegraphics{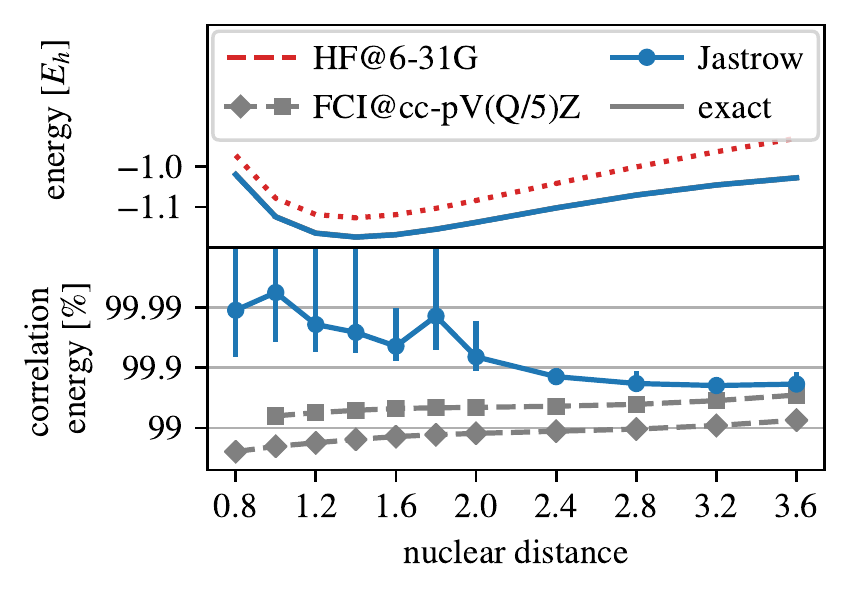}
    \caption{\label{fig:H2_diss}\textbf{Dissociation curve of the hydrogen molecule.} Upper panel shows total energy. Exact results \cite{kol/osPotentialEnergyCurves1965} cannot be distinguished from FCI and the deep Jastrow factor. Lower panel shows the percentage of correlation energy recovered. FCI results were obtained with PySCF~\cite{sunRecentDevelopmentsPySCF2020} in the cc-pVQZ basis (orange) and cc-pV5Z basis (green). Deep QMC surpasses the FCI accuracy for the entire dissociation curve.}
\end{figure}
Next we turn to modelling electron correlation with the deep Jastrow factor (Sec.~\ref{subsec:deep_jastrow}). We start by evaluating the deep Jastrow factor for H$_2$ and He, two-electron closed-shell systems for which the ground state is node-less (the antisymmetry comes from the spin part of the wavefunction only), such that the Jastrow factor is, in principle, sufficient to reproduce exact results. This yields a pure test for the expressiveness of the deep Jastrow factor. The recovered many-body correlation is measured by the correlation energy,
\begin{equation}
    \eta = \frac{E_\text{VMC}-E_\text{HF}}{E_\text{exact}-E_\text{HF}}
\end{equation}
\begin{table}[H]
\centering
\begin{ruledtabular}
\caption{\label{two-electron_systems}\textbf{Results for two-electron node-less systems.}}
\begin{tabular}{lccc}
system & deep Jastrow factor & exact energy & $\eta$ [\%] \\
\hline
H$_2$ ($d=1.4$) & -1.17446(1) & -1.1744748\cite{kol/osPotentialEnergyCurves1965} & 99.97(3) \\
He & 2.90372(1) & -2.9037247\cite{kinoshitaGroundStateHelium1959} & 99.98(2)\\
\end{tabular}
\end{ruledtabular}
\end{table}
For both systems we obtain energies matching five significant digits of the exact references (Table~\ref{two-electron_systems}). We evaluate the ansatz along the dissociation curve of H$_2$ (Fig.~\ref{fig:H2_diss}). Deep QMC outperforms FCI even with the large cc-pV5Z basis set, reducing the error in correlation energy by one to two orders of magnitude at compressed geometries and still being more accurate at stretched geometries, where the system exhibits static correlation and the restricted HF baseline gives qualitatively wrong results (ionic contributions resulting in negative interaction energy). The results demonstrate the difficulty of modelling dynamic correlation in Slater-determinant space when applying purely second-quantized approaches and showcase the advantages of explicitly encoding many-body correlations.

\subsection{Systematically approaching the fixed-node limit}
\begin{figure}[t]
    \centering
    \includegraphics{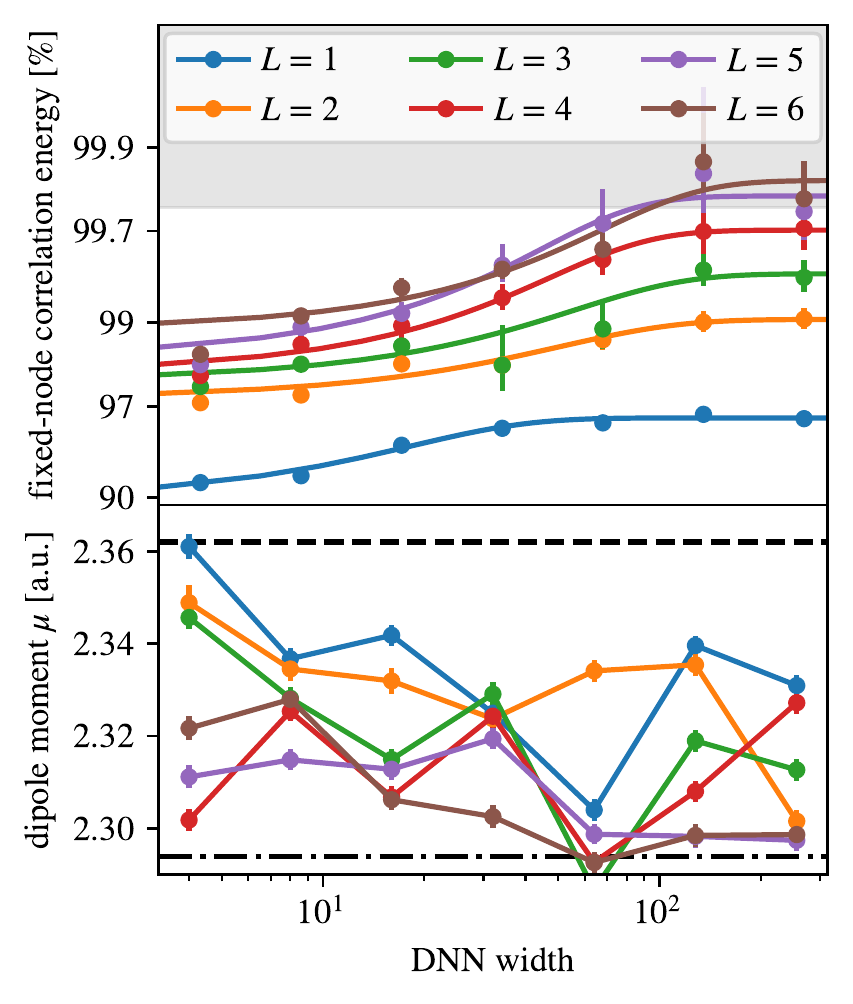}
    \caption{\label{fig:complete_jastrow} \textbf{Approaching FN-DMC accuracy with deep Jastrow factor.} For LiH increasingly expressive Jastrow factors are trained. The dependence on the dimension of the convolutional kernel (DNN width) and the number of interactions ($L$) in the Jastrow graph-neural-network is shown. The orbitals of the antisymmetric part are expressed in the TZP basis\cite{pritchardNewBasisSet2019}. The upper panel shows the energy of the trial wavefunctions. The most accurate ansatzes give results within the sampling error of the FN-DMC energy of a single-determinant benchmark from \citet{casalegnoComputingAccurateForces2003a}, indicated by the shaded region at the upper end of the graph. In the lower panel the dipole moment of the trial wavefunctions is evaluated. As a reference the dipole moment of a HF (dashed line) and of an explicitly-correlated coupled-cluster (CCSD(T)-R12) calculation \cite{tunegaStaticElectricProperties1998} (dashed doted line) are shown. For the models with five and six interactions the dipole moment converges to the results from the coupled-cluster calculation.
    }
\end{figure}
The complexity of modeling correlation increases steeply with the number of particles. We evaluate the performance of the deep Jastrow factor for LiH and the hydrogen rectangle H$_4$. While these four-electron systems exhibit more intricate interactions, they are computationally lightweight, such that the hyperparameter space of the deep ansatzes can be explored exhaustively. With multiple same-spin electrons, the spatial wavefunction is no longer node-less and the single-determinant Slater--Jastrow ansatz possesses a fixed-node error. Instead of comparing to exact energies, we therefore measure the performance of the Jastrow factor with respect to the fixed-node limit estimated from FN-DMC calculations, and reporting the fixed-node correlation energy,
\begin{equation}
    \eta_\text{FN} = \frac{E_\text{VMC}-E_\text{HF}}{E_\text{DMC}-E_\text{HF}}
\end{equation}
As the fixed-node correlation energy is defined for ansatzes with an identical nodal surface, the nodes of the FN-DMC benchmark have to be reconstructed. For the mean-field nodal surface this implies starting from a HF computation with the same basis set.

For the H$_4$ rectangle we performed a scan of all the hyperparameters of the deep Jastrow factor including those of the graph-convolutional neural network architecture (Table~\ref{tab:hyperparameters}). The scan was a grid search that involved the training of 864 models, comprising models with all combinations of the hyperparameters in the vicinity of their default values. In order to reduce the dimensionality of the experiment some hyperparameters were merged and varied together. Further details of the scan can be found in the caption of Fig.~\ref{fig:scan_hyperprameter_H4}, depicting the energies of all the model instances. The experiment aimed at obtaining a first impression of the hyperparameter space and revealed that by increasing the total number of trainable parameters, the fixed-node limit can be approached. The experiment shows that the energy behaves smoothly with respect to changes in the hyperparameters and there are no strong mutual dependencies between hyperparameters. Several important hyperparameters for systematically scaling the architecture can be identified, such as the depth of the neural network $\eta_\mathbf{\theta}$ from \eqref{eq:deep_jastrow}, the number of interactions $L$ and the dimension of the convolutional kernel, referring to the dimension of the latent space where the interactions within the graph-convolutional neural network take place (see \hyperref[appendix]{Appendix}).

The results were used to perform a thorough investigation of the convergence behavior on LiH, varying a subset of hyperparameters and fixing the remaining hyperparameters at suitable values. We show a systematic convergence to the fixed-node limit with an increasing dimension of the convolutional kernel (DNN width) as well as the number of interactions $L$ (Fig.~\ref{fig:complete_jastrow}). This is an indication that the deep Jastrow factor can be extended towards completeness in a computationally feasible way. The remaining fluctuations of the fixed-node correlation energy are caused by the stochasticity of the training and the sampling error of evaluating the energy of the wavefunction. 

\begin{figure}[t]
    \centering
    \includegraphics{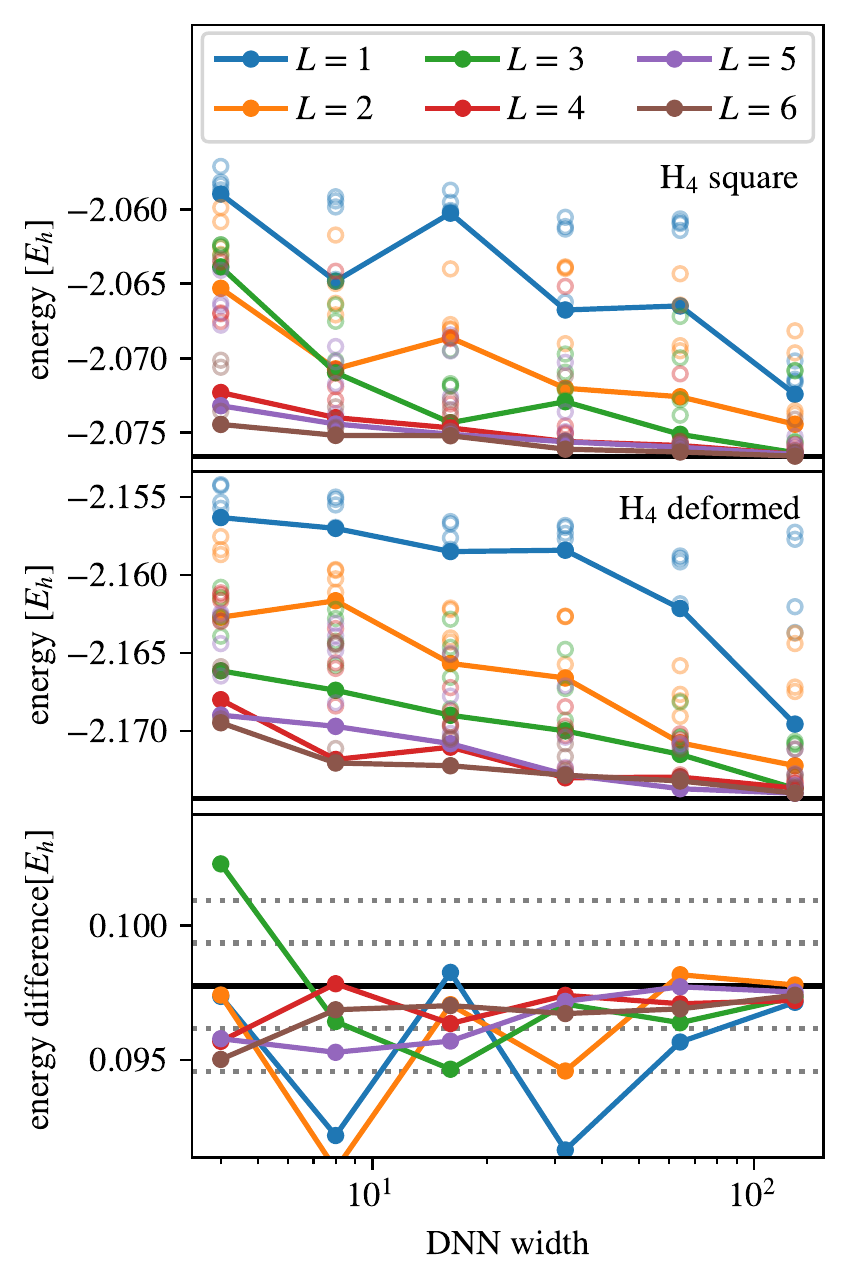}
    \caption{\label{fig:H4_relative} \textbf{Relative energy for hydrogen rectangle.} The convergence of the energy to the DMC reference \cite{gasperichH4ModelSystem2017a} (black line) with increasing model size for two geometries of the hydrogen rectangle, as well as the relative energy difference is shown. The minimal energy over five independent runs is highlighted in the upper plots and used to compute the relative energy. The dashed lines indicate error margins of 1 and 2 kcal/mol.
    }
\end{figure}

By evaluating the dipole moment of the LiH wavefunctions we go beyond the energy and investigate the convergence of a property that the PauliNet ansatz is not explicitly optimized for. We found that upon converging to the fixed-node limit with increasingly large models the dipole moment approaches the coupled cluster reference (Fig.~\ref{fig:complete_jastrow}). Even though the energies of the LiH wavefunctions converge consistently, the convergence of the dipole moment is subject to fluctuations, which are particularly strong for the small models and decrease as the fixed-node limit is approached. This can be explained by degenerate energy minima of the ansatz with respect to the parameters. Multiple solutions to the optimization problem can be present if the exact solution is outside the variational subspace. This ambiguity however decreases with increasing expressiveness of the trial wavefunction.

While the accuracy with respect to the total energy is an appropriate measure for expressiveness of the trial wavefunction ansatz, in practice relative energies are most often of interest. The capability of the full PauliNet ansatz in computing relative energies has been previously demonstrated for the cyclobutadiene automerization\cite{hermannDeepneuralnetworkSolutionElectronic2020} and the results with the deep Jastrow factor for the node-less H$_2$ (Fig. \ref{fig:H2_diss}) provide interaction energies at the level of FCI. Here we want to study how the relative energy converges with an increasing expressiveness of the deep Jastrow factor and demonstrate a cancellation of errors at different geometries. This is a feature that makes relative energy calculations usually more accurate than total energy calculations and is very desirable for any quantum chemistry method. We optimized increasingly expressive versions of the deep Jastrow factor for two geometries of the hydrogen rectangle (Tabel \ref{tab:geometries}) and determined their relative energy (Fig.~\ref{fig:H4_relative}). In order to reduce the level of stochasticity in the training we performed five independent optimization runs and used the ones with the lowest energy to calculate the relative energies. Both the total and relative energy converge to the DMC reference with an increasing number of trainable parameters of the ansatz. Furthermore the relative energy fluctuates within 2 kcal/mol of the DMC reference for all models with more than two interactions, and is well within 1 kcal/mol for the models with the largest DNN width. This demonstrates that the deep Jastrow factor can achieve similar accuracy for both geometries and exhibits a cancellation of errors. The stochasticity of the optimization however complicates the comparison of individual runs, which will be subject to further investigation. Looking at the energy of the different optimizations though, we found a decrease in stochasticity of the final energy with increasing model size, which is convenient for practical purpose, where typically large models would be used. The difficulty of optimizing small models is well-known in the context of training neural networks, which tends to be improved by increasing the number of trainable parameters \cite{livniComputationalEfficiencyTraining}.

\begin{figure}[t]
    \centering
    \includegraphics{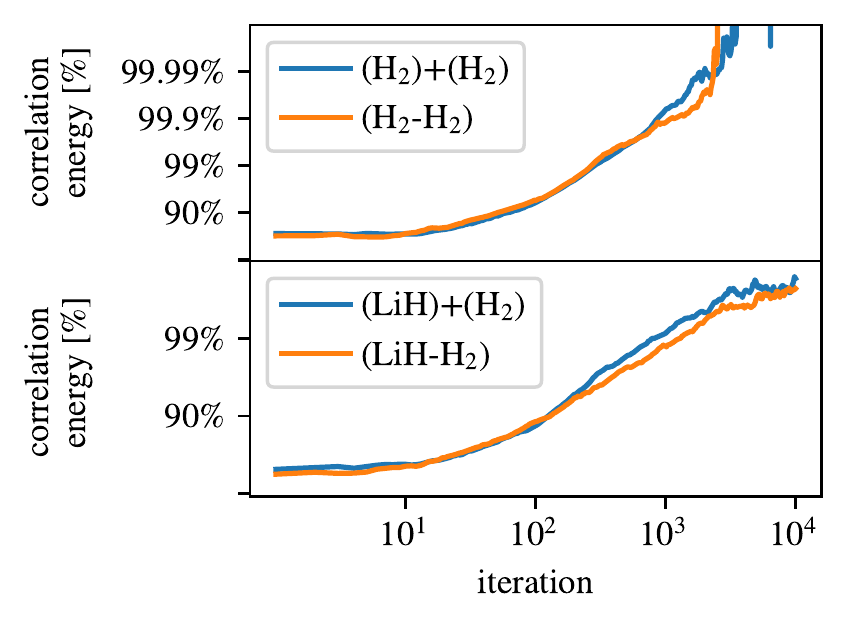}
    \caption{\label{fig:size_consistency} \textbf{Size consistency of deep Jastrow factor optimization.} The figure shows the smoothed training curves of the deep Jastrow factor for two systems of non-interacting molecules. The optimization of an ansatz for the joint system (orange) is compared to the independent optimization of two separate ansatzes for the subsystems respectively (blue).}
\end{figure}

\begin{table}[H]
\centering
\begin{ruledtabular}
\caption{\label{tab:size_consistency}\textbf{Sampled energies of the size consistency experiment.}}
\begin{tabular}{lccc}
system & combined & individual &  exact\cite{kol/osPotentialEnergyCurves1965,cencekBenchmarkCalculationsHe22000} \\
\hline
H$_2$-H$_2$ & -2.34894(1) & -2.34895(1)& -2.34895 \\
LiH-H$_2$ & -9.24394(7) & -9.24405(7) & -9.24501\\
\end{tabular}
\end{ruledtabular}
\end{table}

One of the essential properties of any proper electronic structure method is size consistency.
Traditional Jastrow factors are factorizable in the electronic and nuclear coordinates of two infinitely distant subsystems, which leads to exact size consistency for identical copies of a given system, and to approximate size consistency for an assembly of different systems (because optimized parameters are now shared by different systems).
In PauliNet the embeddings $\mathbf{x}_i$ for two electrons at two distant subsystems are independent of each other by construction.
Although the subsequent nonlinear transformation $\eta_{\boldsymbol\theta}$ applied to the sum of the embeddings breaks exact factorizability, it could be restored by applying the transformation before summing the embeddings, which in numerical experiments does not affect performance.
Regardless, in numerical experiments with two systems of non-interacting molecules (H$_2$-H$_2$ and LiH-H$_2$) we show that even the variant of our ansatz which is not exactly factorizable is size-consistent in practice (Table~\ref{tab:size_consistency}).
For the system composed of two distant hydrogen molecules, both the combined and individual calculations give nearly exact results, 99.99(1)\% and 100.00(1)\% of the correlation energy, respectively.
In the second test with LiH and H$_2$, 99.65(2)\% and 99.68(2)\% of the correlation energy is achieved, respectively, which corresponds to the difference of less than 10\% of the overall error of PauliNet with respect to the exact energy.
The results furthermore show that optimization of the ansatz for the combined system works similarly well as optimizing the separate instances for the respective subsystems (Fig.~\ref{fig:size_consistency}).

\subsection{Application of different levels of theory to H$_2$O}

\begin{figure}[t]
    \centering
    \includegraphics{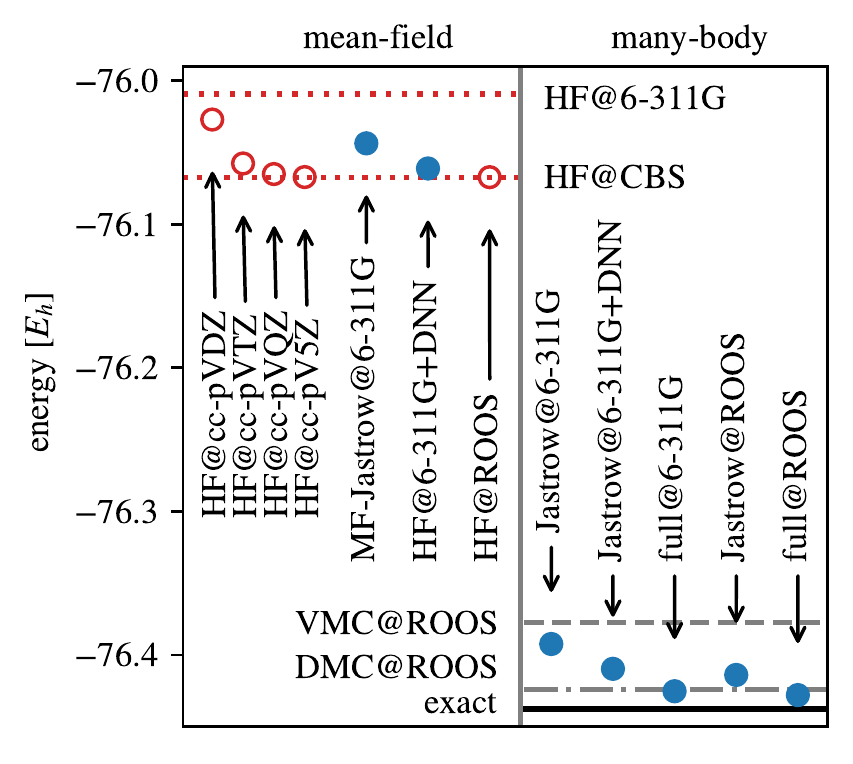}
    \caption{\label{fig:H2O} \textbf{Single-determinant variants of PauliNet evaluated on the H$_2$O molecule.} All three restricted variants of PauliNet introduced in Methods are compared: the deep orbital correction (6-311G+DNN), the full deep Jastrow factor, and the mean-field (MF) Jastrow factor. Furthermore the energy of the combined ansatz (Jastrow@6-311G+DNN) and single-determinant versions of the full PauliNet (full) are shown. ROOS denotes the Roos-aug-DZ-ANO basis set. VMC and DMC references are taken from \citet{gurtubayDissociationEnergyWater2007}. HF@CBS and exact energy are taken from \citet{rosenbergSCFCIStudies1975}. The Monte Carlo sampling error is smaller than the size of the markers. The training curves for the ansatzes are shown in Fig.~\ref{fig:H2O_training}.}
\end{figure}

The results for the small test systems showed that DNNs can be used to converge to the CBS limit within the mean-field theory, and that by adding correlation with a deep Jastrow factor the fixed-node limit can be approached. To investigate how these ansatzes behave for larger systems, we evaluated the respective instances of PauliNet on the water molecule (Fig.~\ref{fig:H2O}, Fig.~\ref{fig:H2O_training}). These experiments aim at demonstrating that the same ansatzes can be applied to a variety of systems without any modifications and test how much their respective accuracy decreases if the size of the graph-convolutional neural network is kept fixed. For the experiment we chose 4 interactions and a kernel dimension of 256, which is equal to the large models from the H$_4$ and LiH experiments. Due to the convolutional nature of the neural network the number of trainable parameters is mostly independent of the number of electrons, hence it is similar to the previous experiments.

\begin{table*}[t]
\caption{\label{H2Oresults}\textbf{Benchmarking single-determinant (SD) Slater--Jastrow (SJ) ansatzes on H$_2$O.}
}
\vspace{2mm}
\centering
\begin{ruledtabular}
\begin{tabular}{lccccc}
reference & HF & VMC (SD-SJ) & DMC (SD-SJ) & $\eta_{FN}$ [\%] & basis set \\
\hline
PauliNet & $-$76.009 & $-$76.3923(7) & -- & 91.2(2)\footnotemark[1] & 6-311G\\
PauliNet & $-$76.0612 & $-$76.4096(7) & -- & 96.0(2)\footnotemark[1] & 6-311G+DNN \\
PauliNet & $-$76.0672 & $-76.4139$(5) & -- & $97.2$(1)\footnotemark[1] & Roos-aug-DZ-ANO \\
\citet{clarkComputingEnergyWater2011}& -- & $-$76.3938(4) &  $-$76.4236(2) & 91.6(1) & Roos-aug-TZ-ANO \\
\citet{gurtubayDissociationEnergyWater2007}& $-$76.0672 & $-$76.3773(2) &  $-$76.42376(5) & 87.01(6) & Roos-aug-DZ-ANO \\
\citet{gurtubayQuantumMonteCarlo2006}& $-$76.0587 & $-$76.327(1)&$-$76.42102(4)& 73.5(3) &  6–311++G(2d,2p)\\
\end{tabular}
\end{ruledtabular}
\footnotesize
\begin{minipage}{0.9\linewidth}
\footnotetext[1]{The fixed-node correlation energy is computed with respect to the reference FN-DMC energy of \citet{gurtubayDissociationEnergyWater2007}.}
\end{minipage}
\end{table*}

We again start with the mean-field theory and consider the finite-basis-set error. We corrected the HF orbitals in the small 6-311G basis set with the deep orbital correction (Sec.~\ref{subsec:deep_orbital}), which recovered 90\% of the finite-basis-set error. We then estimated how much of the finite-basis-set error amounts to the fixed-node error by applying the mean-field Jastrow factor (Sec.~\ref{subsec:mean-field_jastrow}), which can recover about half of the finite-basis-set error only. This suggests that upon approaching the finite-basis-set limit the nodal surface is altered significantly.

Next, we investigate single-determinant ansatzes with the full Jastrow factor (Sec.~\ref{subsec:deep_jastrow}). We benchmark the deep Jastrow factor with a HF determinant in the Roos augmented double-zeta basis (Roos-aug-DZ-ANO)\cite{widmarkDensityMatrixAveraged1990}, a basis set that is frequently used for calculations on H$_2$O and gives HF energies at the CBS limit. We compare to VMC and DMC results from the literature, achieving 97.2(1)\% of the fixed-node correlation energy and surpassing the accuracy of previous VMC calculations with single-determinant Slater--Jastrow trial wavefunctions by half an order of magnitude (Table~\ref{H2Oresults}).

In order to study how finite-basis-set errors manifest in the mean-field nodal surface of both the HF and the many-body ansatzes, we computed the energies of the deep Jastrow factor with a HF determinant in a 6-311G basis. The results suggest that finite-basis-set errors of the HF calculations transfers directly to the many-body regime. In particular, the differences between the energies of the mean-field ansatzes match the differences of the respective Slater--Jastrow trial wavefunctions and errors in the energy due to finite-basis-set effects are not altered by the many-body correlation.

We furthermore demonstrate that both methods can be combined, by optimizing a trial wavefunction composed of a deep Jastrow factor and a Slater determinant of orbitals of an imprecise HF baseline that are modified by the orbital correction. The parameters of both Jastrow factor and orbital correction were optimized simultaneously. The HF baseline was computed in the small 6-311G basis set. With this setup we were able to achieve energies close to the fixed-node limit of the optimal mean-field nodal surface. Starting from a minimal baseline we recovered 96.0(2)\% of the fixed-node correlation energy with respect to the Roos-aug-DZ-ANO basis.

\begin{table}[t]
\centering
\begin{ruledtabular}
\caption{\label{multi-determinant water}\textbf{Comparison of full PauliNet instance with traditional trial wavefunctions.}}
\begin{tabular}{lccc}
 & ansatz &\# determinants & VMC  \\
\hline
PauliNet & SD-SJB & 1 & $-$76.4281(3)   \\
\citet{gurtubayDissociationEnergyWater2007} & SD-SJB  & 1 & $-$76.4034(2) \\
\citet{clarkComputingEnergyWater2011} & MD-SJ & 2316 & $-$76.4259(6) \\
\citet{clarkComputingEnergyWater2011} & MD-SJ &7425 & $-$76.4289(8) \\
\end{tabular}
\end{ruledtabular}
\end{table}

Finally we show that the full PauliNet ansatz can go beyond the fixed-node approximation and train an instance with the same graph-convolutional architecture as in the previous experiments, but using the full backflow transformation. With this ansatz we obtained a VMC energy of -76.4252(3) and -76.4281(3) for the 6-311G and the Roos-aug-DZ-ANO basis set respectively, amounting to 96.67(8)\% and 97.38(8)\% of the total correlation energy. This energy is significantly below the single-determinant DMC results (Fig. \ref{fig:H2O}), demonstrating energetically favorable changes in the nodal surface due to the backflow transformation. A comparison to traditional VMC results shows that the single-determinant version of PauliNet strongly improves on single-determinant Slater--Jastrow--backflow (SD-SJB) trial wavefunctions and  multi-determinant Slater--Jastrow (MD-SJ) wavefunctions need thousands of determinants to obtain a similar accuracy (Table~\ref{multi-determinant water}). Here it should be stated that in principle the accuracy of PauliNet can be further improved by increasing the size of the graph-convolutional neural network architecture or introducing multiple determinants. The comparison should therefore not be understood as ultimate, but serves to give an impression of the capabilities of the PauliNet backflow. More exemplary calculations with the full PauliNet ansatz including multi-determinant ansatzes have been carried out previously\cite{hermannDeepneuralnetworkSolutionElectronic2020} and a more thorough investigation of the improvements in the nodal-surface as well as a benchmark of the computational complexity will be conducted in future work.\\

\section{\label{sec:Discussion}Discussion}

We have demonstrated that the choice of the architecture does not introduce fundamental limitations regarding the flexibility of the investigated components of the PauliNet and that a systematic improvement of the accuracy is possible when increasing the number of trainable parameters in a suitable way. For both the deep orbital correction and the deep Jastrow factor, close to exact energies for the corresponding level of theory can be obtained. This highlights the generality and expressiveness of deep QMC---a single ansatz without any problem-specific modifications can be applied to a variety of systems and extended systematically to improve the accuracy without introducing new components to the trial wavefunction architecture. Though the results with the deep orbital correction and the deep Jastrow factor emphasize the potential of the deep QMC approach, the major benefit of deep QMC over FN-DMC calculations remains that it can go beyond the fixed-node approximation by faithfully representing the nodal surface upon introducing many-body correlation at the level of the orbitals. We have outlined this with an exemplary calculation on the water molecule, using a single-determinant instance of the full PauliNet ansatz. The presented analysis paves the way for future investigations on how the full PauliNet ansatz improves the nodes and overcomes the fixed-node limitations.
\newpage
\section*{Data availability}
The data that support the findings of this study are openly available at \url{http://doi.org/10.6084/m9.figshare.13077158.v3}.
\begin{acknowledgments}
We gratefully acknowledge funding and support from the European Research Commission (ERC CoG 772230), the Berlin Mathematics Research Center MATH+ (Projects AA2-8, EF1-2, AA2-22) and the German Ministry for Education and Research (Berlin Institute for the Foundations of Learning and Data BIFOLD). J. H. would like to thank K.-R. Müller for support and acknowledge funding from TU Berlin. 
\end{acknowledgments}
\section*{References}
\nocite{}
\bibliography{main}

\begin{thebibliography}{36}%
\makeatletter
\providecommand \@ifxundefined [1]{%
 \@ifx{#1\undefined}
}%
\providecommand \@ifnum [1]{%
 \ifnum #1\expandafter \@firstoftwo
 \else \expandafter \@secondoftwo
 \fi
}%
\providecommand \@ifx [1]{%
 \ifx #1\expandafter \@firstoftwo
 \else \expandafter \@secondoftwo
 \fi
}%
\providecommand \natexlab [1]{#1}%
\providecommand \enquote  [1]{``#1''}%
\providecommand \bibnamefont  [1]{#1}%
\providecommand \bibfnamefont [1]{#1}%
\providecommand \citenamefont [1]{#1}%
\providecommand \href@noop [0]{\@secondoftwo}%
\providecommand \href [0]{\begingroup \@sanitize@url \@href}%
\providecommand \@href[1]{\@@startlink{#1}\@@href}%
\providecommand \@@href[1]{\endgroup#1\@@endlink}%
\providecommand \@sanitize@url [0]{\catcode `\\12\catcode `\$12\catcode
  `\&12\catcode `\#12\catcode `\^12\catcode `\_12\catcode `\%12\relax}%
\providecommand \@@startlink[1]{}%
\providecommand \@@endlink[0]{}%
\providecommand \url  [0]{\begingroup\@sanitize@url \@url }%
\providecommand \@url [1]{\endgroup\@href {#1}{\urlprefix }}%
\providecommand \urlprefix  [0]{URL }%
\providecommand \Eprint [0]{\href }%
\providecommand \doibase [0]{http://dx.doi.org/}%
\providecommand \selectlanguage [0]{\@gobble}%
\providecommand \bibinfo  [0]{\@secondoftwo}%
\providecommand \bibfield  [0]{\@secondoftwo}%
\providecommand \translation [1]{[#1]}%
\providecommand \BibitemOpen [0]{}%
\providecommand \bibitemStop [0]{}%
\providecommand \bibitemNoStop [0]{.\EOS\space}%
\providecommand \EOS [0]{\spacefactor3000\relax}%
\providecommand \BibitemShut  [1]{\csname bibitem#1\endcsname}%
\let\auto@bib@innerbib\@empty
\bibitem [{\citenamefont {Foulkes}\ \emph {et~al.}(2001)\citenamefont
  {Foulkes}, \citenamefont {Mitas}, \citenamefont {Needs},\ and\ \citenamefont
  {Rajagopal}}]{foulkesQuantumMonteCarlo2001a}%
  \BibitemOpen
  \bibfield  {author} {\bibinfo {author} {\bibfnamefont {W.~M.~C.}\
  \bibnamefont {Foulkes}}, \bibinfo {author} {\bibfnamefont {L.}~\bibnamefont
  {Mitas}}, \bibinfo {author} {\bibfnamefont {R.~J.}\ \bibnamefont {Needs}}, \
  and\ \bibinfo {author} {\bibfnamefont {G.}~\bibnamefont {Rajagopal}},\ }\href
  {\doibase 10.1103/RevModPhys.73.33} {\bibfield  {journal} {\bibinfo
  {journal} {Rev. Mod. Phys.}\ }\textbf {\bibinfo {volume} {73}},\ \bibinfo
  {pages} {33} (\bibinfo {year} {2001})}\BibitemShut {NoStop}%
\bibitem [{\citenamefont {Sch{\"u}tt}\ \emph {et~al.}(2019)\citenamefont
  {Sch{\"u}tt}, \citenamefont {Gastegger}, \citenamefont {Tkatchenko},
  \citenamefont {M{\"u}ller},\ and\ \citenamefont {Maurer}}]{SchuttNC19}%
  \BibitemOpen
  \bibfield  {author} {\bibinfo {author} {\bibfnamefont {K.~T.}\ \bibnamefont
  {Sch{\"u}tt}}, \bibinfo {author} {\bibfnamefont {M.}~\bibnamefont
  {Gastegger}}, \bibinfo {author} {\bibfnamefont {A.}~\bibnamefont
  {Tkatchenko}}, \bibinfo {author} {\bibfnamefont {K.-R.}\ \bibnamefont
  {M{\"u}ller}}, \ and\ \bibinfo {author} {\bibfnamefont {R.~J.}\ \bibnamefont
  {Maurer}},\ }\href {\doibase 10.1038/s41467-019-12875-2} {\bibfield
  {journal} {\bibinfo  {journal} {Nat. Commun.}\ }\textbf {\bibinfo {volume}
  {10}},\ \bibinfo {pages} {5024} (\bibinfo {year} {2019})}\BibitemShut
  {NoStop}%
\bibitem [{\citenamefont {Gastegger}\ \emph {et~al.}(2020)\citenamefont
  {Gastegger}, \citenamefont {McSloy}, \citenamefont {Luya}, \citenamefont
  {Sch{\"u}tt},\ and\ \citenamefont {Maurer}}]{GasteggerJCP20}%
  \BibitemOpen
  \bibfield  {author} {\bibinfo {author} {\bibfnamefont {M.}~\bibnamefont
  {Gastegger}}, \bibinfo {author} {\bibfnamefont {A.}~\bibnamefont {McSloy}},
  \bibinfo {author} {\bibfnamefont {M.}~\bibnamefont {Luya}}, \bibinfo {author}
  {\bibfnamefont {K.~T.}\ \bibnamefont {Sch{\"u}tt}}, \ and\ \bibinfo {author}
  {\bibfnamefont {R.~J.}\ \bibnamefont {Maurer}},\ }\href {\doibase
  10.1063/5.0012911} {\bibfield  {journal} {\bibinfo  {journal} {J. Chem.
  Phys.}\ }\textbf {\bibinfo {volume} {153}},\ \bibinfo {pages} {044123}
  (\bibinfo {year} {2020})}\BibitemShut {NoStop}%
\bibitem [{\citenamefont {Carleo}\ and\ \citenamefont
  {Troyer}(2017)}]{carleoSolvingQuantumManybody2017}%
  \BibitemOpen
  \bibfield  {author} {\bibinfo {author} {\bibfnamefont {G.}~\bibnamefont
  {Carleo}}\ and\ \bibinfo {author} {\bibfnamefont {M.}~\bibnamefont
  {Troyer}},\ }\href {\doibase 10.1126/science.aag2302} {\bibfield  {journal}
  {\bibinfo  {journal} {Science}\ }\textbf {\bibinfo {volume} {355}},\ \bibinfo
  {pages} {602} (\bibinfo {year} {2017})}\BibitemShut {NoStop}%
\bibitem [{\citenamefont {Choo}, \citenamefont {Mezzacapo},\ and\ \citenamefont
  {Carleo}(2020)}]{chooFermionicNeuralnetworkStates2020a}%
  \BibitemOpen
  \bibfield  {author} {\bibinfo {author} {\bibfnamefont {K.}~\bibnamefont
  {Choo}}, \bibinfo {author} {\bibfnamefont {A.}~\bibnamefont {Mezzacapo}}, \
  and\ \bibinfo {author} {\bibfnamefont {G.}~\bibnamefont {Carleo}},\ }\href
  {\doibase 10.1038/s41467-020-15724-9} {\bibfield  {journal} {\bibinfo
  {journal} {Nature Communications}\ }\textbf {\bibinfo {volume} {11}},\
  \bibinfo {pages} {2368} (\bibinfo {year} {2020})}\BibitemShut {NoStop}%
\bibitem [{\citenamefont {Han}, \citenamefont {Zhang},\ and\ \citenamefont
  {E}(2019)}]{hanSolvingManyelectronSchrodinger2019a}%
  \BibitemOpen
  \bibfield  {author} {\bibinfo {author} {\bibfnamefont {J.}~\bibnamefont
  {Han}}, \bibinfo {author} {\bibfnamefont {L.}~\bibnamefont {Zhang}}, \ and\
  \bibinfo {author} {\bibfnamefont {W.}~\bibnamefont {E}},\ }\href {\doibase
  10.1016/j.jcp.2019.108929} {\bibfield  {journal} {\bibinfo  {journal}
  {Journal of Computational Physics}\ }\textbf {\bibinfo {volume} {399}},\
  \bibinfo {pages} {108929} (\bibinfo {year} {2019})}\BibitemShut {NoStop}%
\bibitem [{\citenamefont {Hermann}, \citenamefont {Sch{\"a}tzle},\ and\
  \citenamefont
  {No{\'e}}(2020)}]{hermannDeepneuralnetworkSolutionElectronic2020}%
  \BibitemOpen
  \bibfield  {author} {\bibinfo {author} {\bibfnamefont {J.}~\bibnamefont
  {Hermann}}, \bibinfo {author} {\bibfnamefont {Z.}~\bibnamefont
  {Sch{\"a}tzle}}, \ and\ \bibinfo {author} {\bibfnamefont {F.}~\bibnamefont
  {No{\'e}}},\ }\href {\doibase 10.1038/s41557-020-0544-y} {\bibfield
  {journal} {\bibinfo  {journal} {Nature Chemistry}\ }\textbf {\bibinfo
  {volume} {12}},\ \bibinfo {pages} {891} (\bibinfo {year} {2020})}\BibitemShut
  {NoStop}%
\bibitem [{\citenamefont {Pfau}\ \emph {et~al.}(2020)\citenamefont {Pfau},
  \citenamefont {Spencer}, \citenamefont {Matthews},\ and\ \citenamefont
  {Foulkes}}]{pfauInitioSolutionManyelectron2020}%
  \BibitemOpen
  \bibfield  {author} {\bibinfo {author} {\bibfnamefont {D.}~\bibnamefont
  {Pfau}}, \bibinfo {author} {\bibfnamefont {J.~S.}\ \bibnamefont {Spencer}},
  \bibinfo {author} {\bibfnamefont {A.~G. D.~G.}\ \bibnamefont {Matthews}}, \
  and\ \bibinfo {author} {\bibfnamefont {W.~M.~C.}\ \bibnamefont {Foulkes}},\
  }\href {\doibase 10.1103/PhysRevResearch.2.033429} {\bibfield  {journal}
  {\bibinfo  {journal} {Phys. Rev. Research}\ }\textbf {\bibinfo {volume}
  {2}},\ \bibinfo {pages} {033429} (\bibinfo {year} {2020})}\BibitemShut
  {NoStop}%
\bibitem [{\citenamefont {Adams}\ \emph {et~al.}(2020)\citenamefont {Adams},
  \citenamefont {Carleo}, \citenamefont {Lovato},\ and\ \citenamefont
  {Rocco}}]{Adams20}%
  \BibitemOpen
  \bibfield  {author} {\bibinfo {author} {\bibfnamefont {C.}~\bibnamefont
  {Adams}}, \bibinfo {author} {\bibfnamefont {G.}~\bibnamefont {Carleo}},
  \bibinfo {author} {\bibfnamefont {A.}~\bibnamefont {Lovato}}, \ and\ \bibinfo
  {author} {\bibfnamefont {N.}~\bibnamefont {Rocco}},\ }\href@noop {} {\enquote
  {\bibinfo {title} {Variational {{Monte Carlo}} calculations of {{A}} {$\leq$}
  4 nuclei with an artificial neural-network correlator ansatz},}\ } (\bibinfo
  {year} {2020}),\ \bibinfo {note} {preprint at
  http://arxiv.org/abs/2007.14282}\BibitemShut {NoStop}%
\bibitem [{\citenamefont {Han}\ \emph {et~al.}(2019)\citenamefont {Han},
  \citenamefont {Li}, \citenamefont {Lin}, \citenamefont {Lu}, \citenamefont
  {Zhang},\ and\ \citenamefont {Zhang}}]{Han19}%
  \BibitemOpen
  \bibfield  {author} {\bibinfo {author} {\bibfnamefont {J.}~\bibnamefont
  {Han}}, \bibinfo {author} {\bibfnamefont {Y.}~\bibnamefont {Li}}, \bibinfo
  {author} {\bibfnamefont {L.}~\bibnamefont {Lin}}, \bibinfo {author}
  {\bibfnamefont {J.}~\bibnamefont {Lu}}, \bibinfo {author} {\bibfnamefont
  {J.}~\bibnamefont {Zhang}}, \ and\ \bibinfo {author} {\bibfnamefont
  {L.}~\bibnamefont {Zhang}},\ }\href@noop {} {\enquote {\bibinfo {title}
  {Universal approximation of symmetric and anti-symmetric functions},}\ }
  (\bibinfo {year} {2019}),\ \bibinfo {note} {preprint at
  http://arxiv.org/abs/1912.01765}\BibitemShut {NoStop}%
\bibitem [{\citenamefont {Hutter}(2020)}]{Hutter20}%
  \BibitemOpen
  \bibfield  {author} {\bibinfo {author} {\bibfnamefont {M.}~\bibnamefont
  {Hutter}},\ }\href@noop {} {\enquote {\bibinfo {title} {On {{Representing}}
  ({{Anti}}){{Symmetric Functions}}},}\ } (\bibinfo {year} {2020}),\ \Eprint
  {http://arxiv.org/abs/2007.15298} {arXiv:2007.15298} \BibitemShut {NoStop}%
\bibitem [{\citenamefont {Drummond}, \citenamefont {Towler},\ and\
  \citenamefont {Needs}(2004)}]{drummondJastrowCorrelationFactor2004a}%
  \BibitemOpen
  \bibfield  {author} {\bibinfo {author} {\bibfnamefont {N.~D.}\ \bibnamefont
  {Drummond}}, \bibinfo {author} {\bibfnamefont {M.~D.}\ \bibnamefont
  {Towler}}, \ and\ \bibinfo {author} {\bibfnamefont {R.~J.}\ \bibnamefont
  {Needs}},\ }\href {\doibase 10.1103/PhysRevB.70.235119} {\bibfield  {journal}
  {\bibinfo  {journal} {Phys. Rev. B}\ }\textbf {\bibinfo {volume} {70}},\
  \bibinfo {pages} {235119} (\bibinfo {year} {2004})}\BibitemShut {NoStop}%
\bibitem [{\citenamefont {Needs}\ \emph {et~al.}(2009)\citenamefont {Needs},
  \citenamefont {Towler}, \citenamefont {Drummond},\ and\ \citenamefont
  {R{\'i}os}}]{needsContinuumVariationalDiffusion2009}%
  \BibitemOpen
  \bibfield  {author} {\bibinfo {author} {\bibfnamefont {R.~J.}\ \bibnamefont
  {Needs}}, \bibinfo {author} {\bibfnamefont {M.~D.}\ \bibnamefont {Towler}},
  \bibinfo {author} {\bibfnamefont {N.~D.}\ \bibnamefont {Drummond}}, \ and\
  \bibinfo {author} {\bibfnamefont {P.~L.}\ \bibnamefont {R{\'i}os}},\ }\href
  {\doibase 10.1088/0953-8984/22/2/023201} {\bibfield  {journal} {\bibinfo
  {journal} {J. Phys.: Condens. Matter}\ }\textbf {\bibinfo {volume} {22}},\
  \bibinfo {pages} {023201} (\bibinfo {year} {2009})}\BibitemShut {NoStop}%
\bibitem [{\citenamefont {Barr}, \citenamefont {Gispen},\ and\ \citenamefont
  {Lamacraft}(2020)}]{BarrPMLR20}%
  \BibitemOpen
  \bibfield  {author} {\bibinfo {author} {\bibfnamefont {A.}~\bibnamefont
  {Barr}}, \bibinfo {author} {\bibfnamefont {W.}~\bibnamefont {Gispen}}, \ and\
  \bibinfo {author} {\bibfnamefont {A.}~\bibnamefont {Lamacraft}},\ }in\ \href
  {http://proceedings.mlr.press/v107/barr20a.html} {\emph {\bibinfo {booktitle}
  {Proceedings of {{Machine Learning Research}}}}},\ Vol.\ \bibinfo {volume}
  {107}\ (\bibinfo  {publisher} {{PMLR}},\ \bibinfo {year} {2020})\ pp.\
  \bibinfo {pages} {635--653}\BibitemShut {NoStop}%
\bibitem [{\citenamefont {Han}, \citenamefont {Lu},\ and\ \citenamefont
  {Zhou}(2020)}]{Han20}%
  \BibitemOpen
  \bibfield  {author} {\bibinfo {author} {\bibfnamefont {J.}~\bibnamefont
  {Han}}, \bibinfo {author} {\bibfnamefont {J.}~\bibnamefont {Lu}}, \ and\
  \bibinfo {author} {\bibfnamefont {M.}~\bibnamefont {Zhou}},\ }\href {\doibase
  10.1016/j.jcp.2020.109792} {\bibfield  {journal} {\bibinfo  {journal}
  {Journal of Computational Physics}\ }\textbf {\bibinfo {volume} {423}},\
  \bibinfo {pages} {109792} (\bibinfo {year} {2020})}\BibitemShut {NoStop}%
\bibitem [{\citenamefont {L{\'o}pez~R{\'i}os}\ \emph
  {et~al.}(2006)\citenamefont {L{\'o}pez~R{\'i}os}, \citenamefont {Ma},
  \citenamefont {Drummond}, \citenamefont {Towler},\ and\ \citenamefont
  {Needs}}]{lopezriosInhomogeneousBackflowTransformations2006a}%
  \BibitemOpen
  \bibfield  {author} {\bibinfo {author} {\bibfnamefont {P.}~\bibnamefont
  {L{\'o}pez~R{\'i}os}}, \bibinfo {author} {\bibfnamefont {A.}~\bibnamefont
  {Ma}}, \bibinfo {author} {\bibfnamefont {N.~D.}\ \bibnamefont {Drummond}},
  \bibinfo {author} {\bibfnamefont {M.~D.}\ \bibnamefont {Towler}}, \ and\
  \bibinfo {author} {\bibfnamefont {R.~J.}\ \bibnamefont {Needs}},\ }\href
  {\doibase 10.1103/PhysRevE.74.066701} {\bibfield  {journal} {\bibinfo
  {journal} {Phys. Rev. E}\ }\textbf {\bibinfo {volume} {74}},\ \bibinfo
  {pages} {066701} (\bibinfo {year} {2006})}\BibitemShut {NoStop}%
\bibitem [{\citenamefont
  {Kato}(1957)}]{katoEigenfunctionsManyparticleSystems1957a}%
  \BibitemOpen
  \bibfield  {author} {\bibinfo {author} {\bibfnamefont {T.}~\bibnamefont
  {Kato}},\ }\href {\doibase 10.1002/cpa.3160100201} {\bibfield  {journal}
  {\bibinfo  {journal} {Communications on Pure and Applied Mathematics}\
  }\textbf {\bibinfo {volume} {10}},\ \bibinfo {pages} {151} (\bibinfo {year}
  {1957})}\BibitemShut {NoStop}%
\bibitem [{\citenamefont {Hermann}, \citenamefont {Sch{\"a}tzle},\ and\
  \citenamefont {Sauceda}(2021)}]{janhermannDeepqmcDeepqmcDeepQMC2021}%
  \BibitemOpen
  \bibfield  {author} {\bibinfo {author} {\bibfnamefont {J.}~\bibnamefont
  {Hermann}}, \bibinfo {author} {\bibfnamefont {Z.}~\bibnamefont
  {Sch{\"a}tzle}}, \ and\ \bibinfo {author} {\bibfnamefont {H.~E.}\
  \bibnamefont {Sauceda}},\ }\href {\doibase 10.5281/zenodo.4473001} {\enquote
  {\bibinfo {title} {Deepqmc/deepqmc: {{DeepQMC}} 0.3.0},}\ }\bibinfo
  {howpublished} {Zenodo} (\bibinfo {year} {2021})\BibitemShut {NoStop}%
\bibitem [{\citenamefont {Ma}\ \emph {et~al.}(2005)\citenamefont {Ma},
  \citenamefont {Towler}, \citenamefont {Drummond},\ and\ \citenamefont
  {Needs}}]{maSchemeAddingElectron2005a}%
  \BibitemOpen
  \bibfield  {author} {\bibinfo {author} {\bibfnamefont {A.}~\bibnamefont
  {Ma}}, \bibinfo {author} {\bibfnamefont {M.~D.}\ \bibnamefont {Towler}},
  \bibinfo {author} {\bibfnamefont {N.~D.}\ \bibnamefont {Drummond}}, \ and\
  \bibinfo {author} {\bibfnamefont {R.~J.}\ \bibnamefont {Needs}},\ }\href
  {\doibase 10.1063/1.1940588} {\bibfield  {journal} {\bibinfo  {journal} {J.
  Chem. Phys.}\ }\textbf {\bibinfo {volume} {122}},\ \bibinfo {pages} {224322}
  (\bibinfo {year} {2005})}\BibitemShut {NoStop}%
\bibitem [{\citenamefont {Ko{\l{}}os}\ and\ \citenamefont
  {Wolniewicz}(1965)}]{kol/osPotentialEnergyCurves1965}%
  \BibitemOpen
  \bibfield  {author} {\bibinfo {author} {\bibfnamefont {W.}~\bibnamefont
  {Ko{\l{}}os}}\ and\ \bibinfo {author} {\bibfnamefont {L.}~\bibnamefont
  {Wolniewicz}},\ }\href {\doibase 10.1063/1.1697142} {\bibfield  {journal}
  {\bibinfo  {journal} {J. Chem. Phys.}\ }\textbf {\bibinfo {volume} {43}},\
  \bibinfo {pages} {2429} (\bibinfo {year} {1965})}\BibitemShut {NoStop}%
\bibitem [{\citenamefont {Sun}\ \emph {et~al.}(2020)\citenamefont {Sun},
  \citenamefont {Zhang}, \citenamefont {Banerjee}, \citenamefont {Bao},
  \citenamefont {Barbry}, \citenamefont {Blunt}, \citenamefont {Bogdanov},
  \citenamefont {Booth}, \citenamefont {Chen}, \citenamefont {Cui},
  \citenamefont {Eriksen}, \citenamefont {Gao}, \citenamefont {Guo},
  \citenamefont {Hermann}, \citenamefont {Hermes}, \citenamefont {Koh},
  \citenamefont {Koval}, \citenamefont {Lehtola}, \citenamefont {Li},
  \citenamefont {Liu}, \citenamefont {Mardirossian}, \citenamefont {McClain},
  \citenamefont {Motta}, \citenamefont {Mussard}, \citenamefont {Pham},
  \citenamefont {Pulkin}, \citenamefont {Purwanto}, \citenamefont {Robinson},
  \citenamefont {Ronca}, \citenamefont {Sayfutyarova}, \citenamefont
  {Scheurer}, \citenamefont {Schurkus}, \citenamefont {Smith}, \citenamefont
  {Sun}, \citenamefont {Sun}, \citenamefont {Upadhyay}, \citenamefont {Wagner},
  \citenamefont {Wang}, \citenamefont {White}, \citenamefont {Whitfield},
  \citenamefont {Williamson}, \citenamefont {Wouters}, \citenamefont {Yang},
  \citenamefont {Yu}, \citenamefont {Zhu}, \citenamefont {Berkelbach},
  \citenamefont {Sharma}, \citenamefont {Sokolov},\ and\ \citenamefont
  {Chan}}]{sunRecentDevelopmentsPySCF2020}%
  \BibitemOpen
  \bibfield  {author} {\bibinfo {author} {\bibfnamefont {Q.}~\bibnamefont
  {Sun}}, \bibinfo {author} {\bibfnamefont {X.}~\bibnamefont {Zhang}}, \bibinfo
  {author} {\bibfnamefont {S.}~\bibnamefont {Banerjee}}, \bibinfo {author}
  {\bibfnamefont {P.}~\bibnamefont {Bao}}, \bibinfo {author} {\bibfnamefont
  {M.}~\bibnamefont {Barbry}}, \bibinfo {author} {\bibfnamefont {N.~S.}\
  \bibnamefont {Blunt}}, \bibinfo {author} {\bibfnamefont {N.~A.}\ \bibnamefont
  {Bogdanov}}, \bibinfo {author} {\bibfnamefont {G.~H.}\ \bibnamefont {Booth}},
  \bibinfo {author} {\bibfnamefont {J.}~\bibnamefont {Chen}}, \bibinfo {author}
  {\bibfnamefont {Z.-H.}\ \bibnamefont {Cui}}, \bibinfo {author} {\bibfnamefont
  {J.~J.}\ \bibnamefont {Eriksen}}, \bibinfo {author} {\bibfnamefont
  {Y.}~\bibnamefont {Gao}}, \bibinfo {author} {\bibfnamefont {S.}~\bibnamefont
  {Guo}}, \bibinfo {author} {\bibfnamefont {J.}~\bibnamefont {Hermann}},
  \bibinfo {author} {\bibfnamefont {M.~R.}\ \bibnamefont {Hermes}}, \bibinfo
  {author} {\bibfnamefont {K.}~\bibnamefont {Koh}}, \bibinfo {author}
  {\bibfnamefont {P.}~\bibnamefont {Koval}}, \bibinfo {author} {\bibfnamefont
  {S.}~\bibnamefont {Lehtola}}, \bibinfo {author} {\bibfnamefont
  {Z.}~\bibnamefont {Li}}, \bibinfo {author} {\bibfnamefont {J.}~\bibnamefont
  {Liu}}, \bibinfo {author} {\bibfnamefont {N.}~\bibnamefont {Mardirossian}},
  \bibinfo {author} {\bibfnamefont {J.~D.}\ \bibnamefont {McClain}}, \bibinfo
  {author} {\bibfnamefont {M.}~\bibnamefont {Motta}}, \bibinfo {author}
  {\bibfnamefont {B.}~\bibnamefont {Mussard}}, \bibinfo {author} {\bibfnamefont
  {H.~Q.}\ \bibnamefont {Pham}}, \bibinfo {author} {\bibfnamefont
  {A.}~\bibnamefont {Pulkin}}, \bibinfo {author} {\bibfnamefont
  {W.}~\bibnamefont {Purwanto}}, \bibinfo {author} {\bibfnamefont {P.~J.}\
  \bibnamefont {Robinson}}, \bibinfo {author} {\bibfnamefont {E.}~\bibnamefont
  {Ronca}}, \bibinfo {author} {\bibfnamefont {E.~R.}\ \bibnamefont
  {Sayfutyarova}}, \bibinfo {author} {\bibfnamefont {M.}~\bibnamefont
  {Scheurer}}, \bibinfo {author} {\bibfnamefont {H.~F.}\ \bibnamefont
  {Schurkus}}, \bibinfo {author} {\bibfnamefont {J.~E.~T.}\ \bibnamefont
  {Smith}}, \bibinfo {author} {\bibfnamefont {C.}~\bibnamefont {Sun}}, \bibinfo
  {author} {\bibfnamefont {S.-N.}\ \bibnamefont {Sun}}, \bibinfo {author}
  {\bibfnamefont {S.}~\bibnamefont {Upadhyay}}, \bibinfo {author}
  {\bibfnamefont {L.~K.}\ \bibnamefont {Wagner}}, \bibinfo {author}
  {\bibfnamefont {X.}~\bibnamefont {Wang}}, \bibinfo {author} {\bibfnamefont
  {A.}~\bibnamefont {White}}, \bibinfo {author} {\bibfnamefont {J.~D.}\
  \bibnamefont {Whitfield}}, \bibinfo {author} {\bibfnamefont {M.~J.}\
  \bibnamefont {Williamson}}, \bibinfo {author} {\bibfnamefont
  {S.}~\bibnamefont {Wouters}}, \bibinfo {author} {\bibfnamefont
  {J.}~\bibnamefont {Yang}}, \bibinfo {author} {\bibfnamefont {J.~M.}\
  \bibnamefont {Yu}}, \bibinfo {author} {\bibfnamefont {T.}~\bibnamefont
  {Zhu}}, \bibinfo {author} {\bibfnamefont {T.~C.}\ \bibnamefont {Berkelbach}},
  \bibinfo {author} {\bibfnamefont {S.}~\bibnamefont {Sharma}}, \bibinfo
  {author} {\bibfnamefont {A.~Y.}\ \bibnamefont {Sokolov}}, \ and\ \bibinfo
  {author} {\bibfnamefont {G.~K.-L.}\ \bibnamefont {Chan}},\ }\href {\doibase
  10.1063/5.0006074} {\bibfield  {journal} {\bibinfo  {journal} {J. Chem.
  Phys.}\ }\textbf {\bibinfo {volume} {153}},\ \bibinfo {pages} {024109}
  (\bibinfo {year} {2020})}\BibitemShut {NoStop}%
\bibitem [{\citenamefont {Kinoshita}(1959)}]{kinoshitaGroundStateHelium1959}%
  \BibitemOpen
  \bibfield  {author} {\bibinfo {author} {\bibfnamefont {T.}~\bibnamefont
  {Kinoshita}},\ }\href {\doibase 10.1103/PhysRev.115.366} {\bibfield
  {journal} {\bibinfo  {journal} {Phys. Rev.}\ }\textbf {\bibinfo {volume}
  {115}},\ \bibinfo {pages} {366} (\bibinfo {year} {1959})}\BibitemShut
  {NoStop}%
\bibitem [{\citenamefont {Pritchard}\ \emph {et~al.}(2019)\citenamefont
  {Pritchard}, \citenamefont {Altarawy}, \citenamefont {Didier}, \citenamefont
  {Gibson},\ and\ \citenamefont {Windus}}]{pritchardNewBasisSet2019}%
  \BibitemOpen
  \bibfield  {author} {\bibinfo {author} {\bibfnamefont {B.~P.}\ \bibnamefont
  {Pritchard}}, \bibinfo {author} {\bibfnamefont {D.}~\bibnamefont {Altarawy}},
  \bibinfo {author} {\bibfnamefont {B.}~\bibnamefont {Didier}}, \bibinfo
  {author} {\bibfnamefont {T.~D.}\ \bibnamefont {Gibson}}, \ and\ \bibinfo
  {author} {\bibfnamefont {T.~L.}\ \bibnamefont {Windus}},\ }\href {\doibase
  10.1021/acs.jcim.9b00725} {\bibfield  {journal} {\bibinfo  {journal} {J.
  Chem. Inf. Model.}\ }\textbf {\bibinfo {volume} {59}},\ \bibinfo {pages}
  {4814} (\bibinfo {year} {2019})}\BibitemShut {NoStop}%
\bibitem [{\citenamefont {Casalegno}, \citenamefont {Mella},\ and\
  \citenamefont {Rappe}(2003)}]{casalegnoComputingAccurateForces2003a}%
  \BibitemOpen
  \bibfield  {author} {\bibinfo {author} {\bibfnamefont {M.}~\bibnamefont
  {Casalegno}}, \bibinfo {author} {\bibfnamefont {M.}~\bibnamefont {Mella}}, \
  and\ \bibinfo {author} {\bibfnamefont {A.~M.}\ \bibnamefont {Rappe}},\ }\href
  {\doibase 10.1063/1.1562605} {\bibfield  {journal} {\bibinfo  {journal} {J.
  Chem. Phys.}\ }\textbf {\bibinfo {volume} {118}},\ \bibinfo {pages} {7193}
  (\bibinfo {year} {2003})}\BibitemShut {NoStop}%
\bibitem [{\citenamefont {Tunega}\ and\ \citenamefont
  {Noga}(1998)}]{tunegaStaticElectricProperties1998}%
  \BibitemOpen
  \bibfield  {author} {\bibinfo {author} {\bibfnamefont {D.}~\bibnamefont
  {Tunega}}\ and\ \bibinfo {author} {\bibfnamefont {J.}~\bibnamefont {Noga}},\
  }\href {\doibase 10.1007/s002140050368} {\bibfield  {journal} {\bibinfo
  {journal} {Theor Chem Acc}\ }\textbf {\bibinfo {volume} {100}},\ \bibinfo
  {pages} {78} (\bibinfo {year} {1998})}\BibitemShut {NoStop}%
\bibitem [{\citenamefont {Gasperich}, \citenamefont {Deible},\ and\
  \citenamefont {Jordan}(2017)}]{gasperichH4ModelSystem2017a}%
  \BibitemOpen
  \bibfield  {author} {\bibinfo {author} {\bibfnamefont {K.}~\bibnamefont
  {Gasperich}}, \bibinfo {author} {\bibfnamefont {M.}~\bibnamefont {Deible}}, \
  and\ \bibinfo {author} {\bibfnamefont {K.~D.}\ \bibnamefont {Jordan}},\
  }\href {\doibase 10.1063/1.4986216} {\bibfield  {journal} {\bibinfo
  {journal} {J. Chem. Phys.}\ }\textbf {\bibinfo {volume} {147}},\ \bibinfo
  {pages} {074106} (\bibinfo {year} {2017})}\BibitemShut {NoStop}%
\bibitem [{\citenamefont {Livni}, \citenamefont {{Shalev-Shwartz}},\ and\
  \citenamefont {Shamir}(2014)}]{livniComputationalEfficiencyTraining}%
  \BibitemOpen
  \bibfield  {author} {\bibinfo {author} {\bibfnamefont {R.}~\bibnamefont
  {Livni}}, \bibinfo {author} {\bibfnamefont {S.}~\bibnamefont
  {{Shalev-Shwartz}}}, \ and\ \bibinfo {author} {\bibfnamefont
  {O.}~\bibnamefont {Shamir}},\ }in\ \href@noop {} {\emph {\bibinfo {booktitle}
  {Advances in Neural Information Processing Systems}}},\ Vol.~\bibinfo
  {volume} {27},\ \bibinfo {editor} {edited by\ \bibinfo {editor}
  {\bibfnamefont {Z.}~\bibnamefont {Ghahramani}}, \bibinfo {editor}
  {\bibfnamefont {M.}~\bibnamefont {Welling}}, \bibinfo {editor} {\bibfnamefont
  {C.}~\bibnamefont {Cortes}}, \bibinfo {editor} {\bibfnamefont
  {N.}~\bibnamefont {Lawrence}}, \ and\ \bibinfo {editor} {\bibfnamefont
  {K.~Q.}\ \bibnamefont {Weinberger}}}\ (\bibinfo  {publisher} {{Curran
  Associates, Inc.}},\ \bibinfo {year} {2014})\ pp.\ \bibinfo {pages}
  {855--863}\BibitemShut {NoStop}%
\bibitem [{\citenamefont {Cencek}\ and\ \citenamefont
  {Rychlewski}(2000)}]{cencekBenchmarkCalculationsHe22000}%
  \BibitemOpen
  \bibfield  {author} {\bibinfo {author} {\bibfnamefont {W.}~\bibnamefont
  {Cencek}}\ and\ \bibinfo {author} {\bibfnamefont {J.}~\bibnamefont
  {Rychlewski}},\ }\href {\doibase 10.1016/S0009-2614(00)00303-1} {\bibfield
  {journal} {\bibinfo  {journal} {Chemical Physics Letters}\ }\textbf {\bibinfo
  {volume} {320}},\ \bibinfo {pages} {549} (\bibinfo {year}
  {2000})}\BibitemShut {NoStop}%
\bibitem [{\citenamefont {Clark}\ \emph {et~al.}(2011)\citenamefont {Clark},
  \citenamefont {Morales}, \citenamefont {McMinis}, \citenamefont {Kim},\ and\
  \citenamefont {Scuseria}}]{clarkComputingEnergyWater2011}%
  \BibitemOpen
  \bibfield  {author} {\bibinfo {author} {\bibfnamefont {B.~K.}\ \bibnamefont
  {Clark}}, \bibinfo {author} {\bibfnamefont {M.~A.}\ \bibnamefont {Morales}},
  \bibinfo {author} {\bibfnamefont {J.}~\bibnamefont {McMinis}}, \bibinfo
  {author} {\bibfnamefont {J.}~\bibnamefont {Kim}}, \ and\ \bibinfo {author}
  {\bibfnamefont {G.~E.}\ \bibnamefont {Scuseria}},\ }\href {\doibase
  10.1063/1.3665391} {\bibfield  {journal} {\bibinfo  {journal} {The Journal of
  Chemical Physics}\ }\textbf {\bibinfo {volume} {135}},\ \bibinfo {pages}
  {244105} (\bibinfo {year} {2011})}\BibitemShut {NoStop}%
\bibitem [{\citenamefont {Gurtubay}\ and\ \citenamefont
  {Needs}(2007)}]{gurtubayDissociationEnergyWater2007}%
  \BibitemOpen
  \bibfield  {author} {\bibinfo {author} {\bibfnamefont {I.~G.}\ \bibnamefont
  {Gurtubay}}\ and\ \bibinfo {author} {\bibfnamefont {R.~J.}\ \bibnamefont
  {Needs}},\ }\href {\doibase 10.1063/1.2770711} {\bibfield  {journal}
  {\bibinfo  {journal} {J. Chem. Phys.}\ }\textbf {\bibinfo {volume} {127}},\
  \bibinfo {pages} {124306} (\bibinfo {year} {2007})}\BibitemShut {NoStop}%
\bibitem [{\citenamefont {Gurtubay}\ \emph {et~al.}(2006)\citenamefont
  {Gurtubay}, \citenamefont {Drummond}, \citenamefont {Towler},\ and\
  \citenamefont {Needs}}]{gurtubayQuantumMonteCarlo2006}%
  \BibitemOpen
  \bibfield  {author} {\bibinfo {author} {\bibfnamefont {I.~G.}\ \bibnamefont
  {Gurtubay}}, \bibinfo {author} {\bibfnamefont {N.~D.}\ \bibnamefont
  {Drummond}}, \bibinfo {author} {\bibfnamefont {M.~D.}\ \bibnamefont
  {Towler}}, \ and\ \bibinfo {author} {\bibfnamefont {R.~J.}\ \bibnamefont
  {Needs}},\ }\href {\doibase 10.1063/1.2150818} {\bibfield  {journal}
  {\bibinfo  {journal} {J. Chem. Phys.}\ }\textbf {\bibinfo {volume} {124}},\
  \bibinfo {pages} {024318} (\bibinfo {year} {2006})}\BibitemShut {NoStop}%
\bibitem [{\citenamefont {Rosenberg}\ and\ \citenamefont
  {Shavitt}(1975)}]{rosenbergSCFCIStudies1975}%
  \BibitemOpen
  \bibfield  {author} {\bibinfo {author} {\bibfnamefont {B.~J.}\ \bibnamefont
  {Rosenberg}}\ and\ \bibinfo {author} {\bibfnamefont {I.}~\bibnamefont
  {Shavitt}},\ }\href {\doibase 10.1063/1.431596} {\bibfield  {journal}
  {\bibinfo  {journal} {The Journal of Chemical Physics}\ }\textbf {\bibinfo
  {volume} {63}},\ \bibinfo {pages} {2162} (\bibinfo {year}
  {1975})}\BibitemShut {NoStop}%
\bibitem [{\citenamefont {Widmark}, \citenamefont {Malmqvist},\ and\
  \citenamefont {Roos}(1990)}]{widmarkDensityMatrixAveraged1990}%
  \BibitemOpen
  \bibfield  {author} {\bibinfo {author} {\bibfnamefont {P.-O.}\ \bibnamefont
  {Widmark}}, \bibinfo {author} {\bibfnamefont {P.-{\AA}.}\ \bibnamefont
  {Malmqvist}}, \ and\ \bibinfo {author} {\bibfnamefont {B.~O.}\ \bibnamefont
  {Roos}},\ }\href {\doibase 10.1007/BF01120130} {\bibfield  {journal}
  {\bibinfo  {journal} {Theoret. Chim. Acta}\ }\textbf {\bibinfo {volume}
  {77}},\ \bibinfo {pages} {291} (\bibinfo {year} {1990})}\BibitemShut
  {NoStop}%
\bibitem [{\citenamefont {Battaglia}\ \emph {et~al.}(2018)\citenamefont
  {Battaglia}, \citenamefont {Hamrick}, \citenamefont {Bapst}, \citenamefont
  {{Sanchez-Gonzalez}}, \citenamefont {Zambaldi}, \citenamefont {Malinowski},
  \citenamefont {Tacchetti}, \citenamefont {Raposo}, \citenamefont {Santoro},
  \citenamefont {Faulkner}, \citenamefont {Gulcehre}, \citenamefont {Song},
  \citenamefont {Ballard}, \citenamefont {Gilmer}, \citenamefont {Dahl},
  \citenamefont {Vaswani}, \citenamefont {Allen}, \citenamefont {Nash},
  \citenamefont {Langston}, \citenamefont {Dyer}, \citenamefont {Heess},
  \citenamefont {Wierstra}, \citenamefont {Kohli}, \citenamefont {Botvinick},
  \citenamefont {Vinyals}, \citenamefont {Li},\ and\ \citenamefont
  {Pascanu}}]{battagliaRelationalInductiveBiases2018}%
  \BibitemOpen
  \bibfield  {author} {\bibinfo {author} {\bibfnamefont {P.~W.}\ \bibnamefont
  {Battaglia}}, \bibinfo {author} {\bibfnamefont {J.~B.}\ \bibnamefont
  {Hamrick}}, \bibinfo {author} {\bibfnamefont {V.}~\bibnamefont {Bapst}},
  \bibinfo {author} {\bibfnamefont {A.}~\bibnamefont {{Sanchez-Gonzalez}}},
  \bibinfo {author} {\bibfnamefont {V.}~\bibnamefont {Zambaldi}}, \bibinfo
  {author} {\bibfnamefont {M.}~\bibnamefont {Malinowski}}, \bibinfo {author}
  {\bibfnamefont {A.}~\bibnamefont {Tacchetti}}, \bibinfo {author}
  {\bibfnamefont {D.}~\bibnamefont {Raposo}}, \bibinfo {author} {\bibfnamefont
  {A.}~\bibnamefont {Santoro}}, \bibinfo {author} {\bibfnamefont
  {R.}~\bibnamefont {Faulkner}}, \bibinfo {author} {\bibfnamefont
  {C.}~\bibnamefont {Gulcehre}}, \bibinfo {author} {\bibfnamefont
  {F.}~\bibnamefont {Song}}, \bibinfo {author} {\bibfnamefont {A.}~\bibnamefont
  {Ballard}}, \bibinfo {author} {\bibfnamefont {J.}~\bibnamefont {Gilmer}},
  \bibinfo {author} {\bibfnamefont {G.}~\bibnamefont {Dahl}}, \bibinfo {author}
  {\bibfnamefont {A.}~\bibnamefont {Vaswani}}, \bibinfo {author} {\bibfnamefont
  {K.}~\bibnamefont {Allen}}, \bibinfo {author} {\bibfnamefont
  {C.}~\bibnamefont {Nash}}, \bibinfo {author} {\bibfnamefont {V.}~\bibnamefont
  {Langston}}, \bibinfo {author} {\bibfnamefont {C.}~\bibnamefont {Dyer}},
  \bibinfo {author} {\bibfnamefont {N.}~\bibnamefont {Heess}}, \bibinfo
  {author} {\bibfnamefont {D.}~\bibnamefont {Wierstra}}, \bibinfo {author}
  {\bibfnamefont {P.}~\bibnamefont {Kohli}}, \bibinfo {author} {\bibfnamefont
  {M.}~\bibnamefont {Botvinick}}, \bibinfo {author} {\bibfnamefont
  {O.}~\bibnamefont {Vinyals}}, \bibinfo {author} {\bibfnamefont
  {Y.}~\bibnamefont {Li}}, \ and\ \bibinfo {author} {\bibfnamefont
  {R.}~\bibnamefont {Pascanu}},\ }\href@noop {} {\enquote {\bibinfo {title}
  {Relational inductive biases, deep learning, and graph networks},}\ }
  (\bibinfo {year} {2018}),\ \Eprint {http://arxiv.org/abs/1806.01261}
  {arXiv:1806.01261} \BibitemShut {NoStop}%
\bibitem [{\citenamefont {Gilmer}\ \emph {et~al.}(2017)\citenamefont {Gilmer},
  \citenamefont {Schoenholz}, \citenamefont {Riley}, \citenamefont {Vinyals},\
  and\ \citenamefont {Dahl}}]{gilmerNeuralMessagePassing2017}%
  \BibitemOpen
  \bibfield  {author} {\bibinfo {author} {\bibfnamefont {J.}~\bibnamefont
  {Gilmer}}, \bibinfo {author} {\bibfnamefont {S.~S.}\ \bibnamefont
  {Schoenholz}}, \bibinfo {author} {\bibfnamefont {P.~F.}\ \bibnamefont
  {Riley}}, \bibinfo {author} {\bibfnamefont {O.}~\bibnamefont {Vinyals}}, \
  and\ \bibinfo {author} {\bibfnamefont {G.~E.}\ \bibnamefont {Dahl}},\ }in\
  \href@noop {} {\emph {\bibinfo {booktitle} {International {{Conference}} on
  {{Machine Learning}}}}}\ (\bibinfo  {publisher} {{PMLR}},\ \bibinfo {year}
  {2017})\ pp.\ \bibinfo {pages} {1263--1272}\BibitemShut {NoStop}%
\bibitem [{\citenamefont {Sch{\"u}tt}\ \emph {et~al.}(2018)\citenamefont
  {Sch{\"u}tt}, \citenamefont {Sauceda}, \citenamefont {Kindermans},
  \citenamefont {Tkatchenko},\ and\ \citenamefont
  {M{\"u}ller}}]{schuttSchNetDeepLearning2018a}%
  \BibitemOpen
  \bibfield  {author} {\bibinfo {author} {\bibfnamefont {K.~T.}\ \bibnamefont
  {Sch{\"u}tt}}, \bibinfo {author} {\bibfnamefont {H.~E.}\ \bibnamefont
  {Sauceda}}, \bibinfo {author} {\bibfnamefont {P.-J.}\ \bibnamefont
  {Kindermans}}, \bibinfo {author} {\bibfnamefont {A.}~\bibnamefont
  {Tkatchenko}}, \ and\ \bibinfo {author} {\bibfnamefont {K.-R.}\ \bibnamefont
  {M{\"u}ller}},\ }\href {\doibase 10.1063/1.5019779} {\bibfield  {journal}
  {\bibinfo  {journal} {J. Chem. Phys.}\ }\textbf {\bibinfo {volume} {148}},\
  \bibinfo {pages} {241722} (\bibinfo {year} {2018})}\BibitemShut {NoStop}%
\end{thebibliography}%
\newpage
\section*{Appendix}
\appendix{}\label{appendix}
\subsection*{Graph-convolutional neural-network architecture}
At the core of the PauliNet architecture a graph-convolutional neural network generates a permutation-equivariant latent-space many-body representation of a given electron configuration. In the following we give a short introduction to the network architecture, discussing both the general concept and the particular application in the context of PauliNet.

Graph neural networks are constructed to represent functions on graph domains\cite{battagliaRelationalInductiveBiases2018} and have become increasing popular for modeling chemical systems, as they can be designed to comply with symmetries of molecules\cite{gilmerNeuralMessagePassing2017}. The graph-convolutional neural network of PauliNet is a modification of SchNet \cite{schuttSchNetDeepLearning2018a}, an architecture developed to predict molecular properties from atom positions upon being trained in a supervised setting, that is by repeated exposure to known pairs of input and output data. In SchNet a trainable embedding is assigned to each atom, which serves to give an abstract representation of the atomic properties in a high-dimensional feature space and is successively updated to encode information about the atomic environment. The updates are implemented as convolutions over the graph of atomic distances, which makes the architecture invariant to translation, rotation and equivariant with respect to the exchange of identical atoms. The graph convolutions furthermore implement parameter sharing across the edges, such that the number of network parameters does not depend on the number of interacting entities, hence is constant with system size. The final features are then used to predict the molecular properties respectively.

In quantum chemistry we consider modeling electrons and nuclei. At this level a molecule can be represented as a complete graph, where nodes correspond to electrons and nuclei and the distances of each pair of particles is assigned to the edge between their respective nodes. The graph-convolutional neural network at the core of PauliNet acts on this graph representation of the system. Similar to the SchNet implementation a representation in an abstract feature space is assigned to each node by introducing electronic embeddings $\mathbf X_{\boldsymbol\theta,s_i}$ and nuclear embeddings $\mathbf Y_{\boldsymbol\theta,I}$ respectively. The embeddings are trainable arrays that are initialized randomly. As same-spin electrons are indistinguishable they share representations and get initialized with a copy of the same electronic embedding,
\begin{equation}
    \mathbf x_i^{(0)}:=\mathbf X_{\boldsymbol\theta,s_i}
\end{equation}
The electronic embeddings constitute the latent-space representation, which serves to obtain Jastrow factor and backflow transformation in the later process of evaluating the trial wavefunction. In order to encode positional information of each electron with respect to the nuclei as well as electronic many-body correlation into the latent-space representation, the electronic embeddings are updated in an interaction process. Information is transmitted along the edges of the graph, by exchanging messages that take the distances to the nuclei and the other electrons into account,
\begin{equation}\label{eq:message}
\begin{aligned}
\mathbf z_i^{(n,\mathrm n)}&:=\sum\nolimits_I
  \mathbf w_{\boldsymbol\theta}^{(n,\mathrm n)}
  \big(\mathbf e(\lvert\mathbf r_i-\mathbf R_I\rvert)\big)
  \odot\mathbf Y_{\boldsymbol\theta,I} \\
\mathbf z_i^{(n,\pm)}&:=\sum\nolimits_{j\neq i}^\pm
  \mathbf w^{(n,\pm)}_{\boldsymbol\theta}
  \big(\mathbf e(\lvert\mathbf r_i-\mathbf r_j\rvert)\big)
  \odot\mathbf h_{\boldsymbol\theta}^{(n)}\big(\mathbf x_j^{(n)}\big) \\ 
\end{aligned}
\end{equation}
Here the functions $\mathbf{w_\theta}$ and $\mathbf{h_\theta}$ are implemented by fully-connected neural networks, $\mathbf{e}$ represents an expansion of the distances in a basis of Gaussian functions and $\odot$ indicates element-wise multiplication. For each two interacting particles the filter generating function $\mathbf{w_\theta}$ generates a mask that is applied to their respective embeddings. Thereby the filter-generating function moderates interactions based on the distance of the particles. By summing up the messages of identical particles their overall contribution is invariant under the exchange of these identical particles. The transformation $\mathbf{h_\theta}$ serves to introduce additional flexibility to the architecture, by separating the latent-space representation from the interaction space. The superscripts of the neural networks indicate that different functions are applied at each subsequent interaction and the filter-generating functions for the interactions with spin-up electrons, spin-down electrons and nuclei are different. The distance expansion $\mathbf{e}$ is truncated with an envelope that ensures it to be cusp-less, that is that all Gaussian features have a vanishing derivative at zero distance, and imposes a long-range cutoff. The final step in the interaction process is to update to the electronic embeddings
\begin{equation}\label{eq:update}
\mathbf x_i^{(n+1)}:=\mathbf x_i^{(n)}
  +\sum\nolimits_\pm\mathbf g^{(n,\pm)}_{\boldsymbol\theta}
  \big(\mathbf z_i^{(n,\pm)}\big)
  +\mathbf g^{(n,\mathrm n)}_{\boldsymbol\theta}
  \big(\mathbf z_i^{(n,\mathrm n)}\big)
\end{equation}
Therefore the messages are transformed from the interaction space to the embedding space and added to the original embedding. The transformation $\mathbf{g_\theta}$ is again implemented by fully-connected neural networks. This interaction process is repeated $L$ times, to successively encode increasingly complex many-body information. The continuous-filter convolutions over the molecular graph and the initialization of electrons with identical embeddings make the architecture equivariant with respect to the exchange of same-spin electrons,
\begin{equation}
 \mathcal P_{ij}{\mathbf x_i}(\mathbf r)=\mathbf x_j(\mathcal P_{ij}\mathbf r)
\end{equation}
Overall this gives a latent-space representation that can efficiently encode electronic many-body effects while intrinsically fulfilling the desired permutation equivariance. Information about the hyperparameters of all of the components is collected in Table~\ref{tab:hyperparameters}.

A single-particle variant of the graph-convolutional neural-network architecture can be obtained by considering only interactions along edges between electrons and nuclei. That is the overall architecture of the network remains identical but the electronic updates $\mathbf z_i^{(n,\pm)}$ in \eqref{eq:update} are removed. Given that the convolution over the electronic distances is the only interaction between electrons, the final embeddings do not contain any many-body correlation.
\begin{table*}[t]

\caption{\textbf{Components of the deep Jastrow factor.}
\label{tab:hyperparameters}}
\begin{ruledtabular}
\begin{tabular}{lll}
component & type & hyperparameter \\
\hline
 $\mathbf X_{\boldsymbol\theta,s_i}$ -- electronic embedding & trainable array & embedding dimension \\
$\mathbf Y_{\boldsymbol\theta,I}$ -- nuclear embeddings  & trainable array & kernel dimension  \\
$\mathbf e$ -- distance expansion  & fixed function & \# distance features \\
$\mathbf{w_\theta}$ -- filter generating function  & DNN & (\# distance features $\rightarrow$ kernel dimension), depth\\
$\mathbf{h_\theta}$ -- transformation embedding to kernel space & DNN & (embedding dimension $\rightarrow$ kernel dimension), depth\\
$\mathbf{g_\theta}$ -- transformation kernel to embedding space  & DNN & (kernel dimension $\rightarrow$ embedding dimension), depth\\
$\eta_{\boldsymbol\theta}$ -- Jastrow network & DNN & (embedding dimension $\rightarrow$ 1), depth\\
full architecture & --- & \# interactions $L$
\end{tabular}
\end{ruledtabular}
\end{table*}
\begin{figure}[H]
    \centering
    \includegraphics{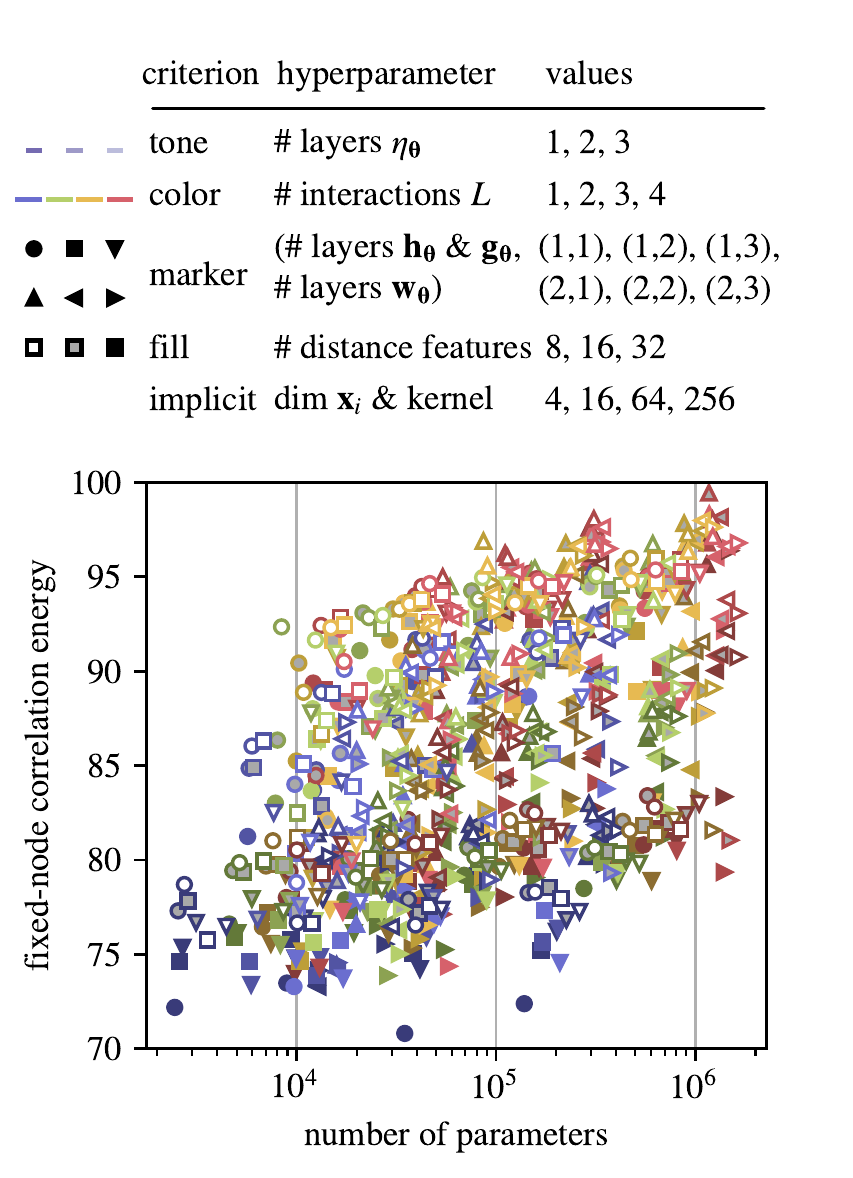}
    \caption{\label{fig:scan_hyperprameter_H4} \textbf{Hyperparameter scan of the H$_4$ square.} The figure maps models with different combinations of hyperparameters in the space of accuracy with respect to the number of total trainable parameters. The exact set of hyperparameters of each model can be decoded via the accompanying legend. In order to reduce the degrees of freedom the embedding and kernel dimensions were joined and varied together. The models with increasing embedding and kernel dimensions can be distinguished by their number of total parameters, hence no explicit criterion is introduced. The energy of the models is estimated from the final steps of the optimization procedure. The fixed-node correlation energy is obtained with respect to FN-DMC results \cite{gasperichH4ModelSystem2017a}, where a single-determinant trial wavefunction with a basis consisting of s functions from the cc-pV5Z basis set and the p and d functions from the cc-pVTZ basis set was used. Increasing the number of trainable parameters the fixed-node limit can be approached. }
\end{figure}

\begin{figure}[H]
    \centering
    \includegraphics{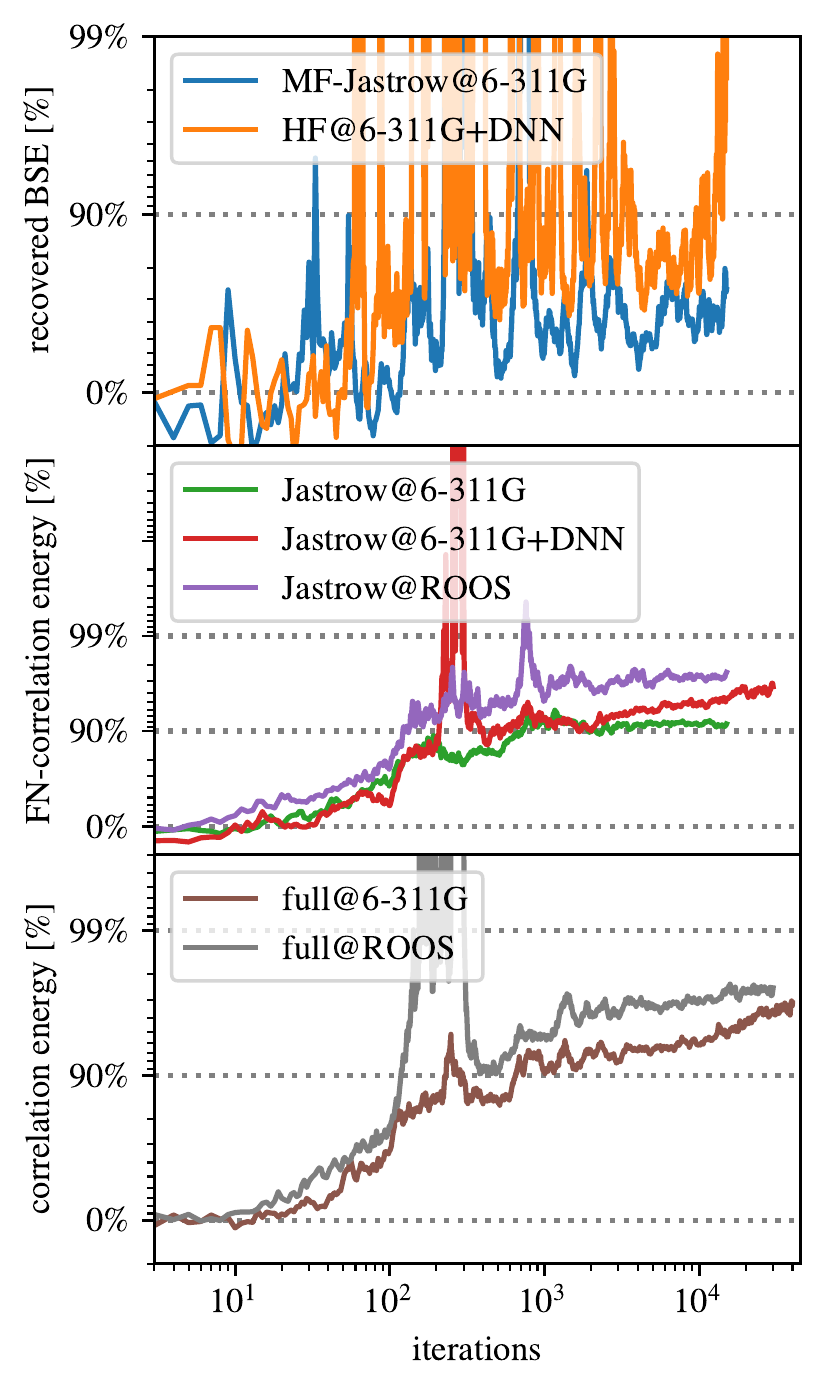}
    \caption{\label{fig:H2O_training} \textbf{Training curve for the H$_2$O experiment.} Exponential walking averages of the energy along the optimization process are shown for the ansatzes from Fig.~\ref{fig:H2O}. Upper panel depicts the percentage of the recovered finite-basis-set error (BSE) for the mean-field ansatzes, given by the energy difference of the HF@6-311G baseline from the complete-basis-set limit estimate. The center panel gives the fixed-node correlation energy of the Slater--Jastrow type versions of PauliNet with respect to the DMC reference in the Roos-aug-DZ-ANO basis set \cite{gurtubayDissociationEnergyWater2007}. Lower panel shows the total correlation energy for the full PauliNet ansatzes.}
\end{figure}

\begin{table}[H]
\caption{\textbf{Hyperparameters used in numerical calculations.}
\label{tab:hyperparameters_training}}
\centering
\begin{ruledtabular}
\begin{tabular}{ll}
Hyperparameter & Value \\
\hline
One-electron basis & 6-31G \\  
Dimension of $\mathbf e$ (\# distance features) & 16 \\
Dimension of $\mathbf x_i$ (embedding dimension)& 128\\
Number of interaction layers $L$ & 4 \\
Number of layers in $\eta_{\boldsymbol\theta}$ & 3 \\
Number of layers in $\mathbf w_{\boldsymbol\theta}$ & 1 \\
Number of layers in $\mathbf h_{\boldsymbol\theta}$ & 2 \\
Number of layers in $\mathbf g_{\boldsymbol\theta}$ & 2 \\
Batch size & {2000} \\
Number of walkers & {2000} \\
\multirow{6}{*}{Number of training steps}&H$_4$:~{5000}\\
    &H$_2$:~{10000} \\
    &He:~{10000} \\
    &Be:~{10000} \\
    &LiH:~{10000} \\
    & H$_2$O:~see Fig.~\ref{fig:H2O_training} \\
Optimizer & AdamW \\
Learning rate scheduler & CyclicLR \\
Minimum/maximum learning rate & 0.0001/0.01 \\
Clipping window $q$ & 5 \\
Epoch size & 100 \\
Number of decorrelation sampling steps & 4 \\
Target acceptance & 57\% \\
\end{tabular}
\end{ruledtabular}
\end{table}

\begin{table}[h!]
\caption{\textbf{Geometries of test systems.}\label{tab:geometries}}
\centering
\begin{ruledtabular}
\begin{tabular}{llc}
Molecule & Atom & position [\AA]\\
\hline
\multirow{2}{*}{LiH}&Li&(0.000, 0.000, 0.000)\\
    &H&(1.595, 0.000, 0.000) \\
\hline
\multirow{4}{*}{H$_4$ square}&H&(-0.635, -0.635, 0.000)\\
    &H&(-0.635, 0.635, 0.000) \\
    &H&(0.635, -0.635, 0.000) \\
    &H&(0.635, 0.635, 0.000) \\
\hline
\multirow{4}{*}{H$_4$ deformed}&H&(-0.900, -0.635, 0.000)\\
    &H&(-0.900, 0.635, 0.000) \\
    &H&(0.900, -0.635, 0.000) \\
    &H&(0.900, 0.635, 0.000) \\
\hline
\multirow{3}{*}{H$_2$O}&O&(0.00000, 0.00000, 0.00000)\\
    &H&(0.75695, 0.58588, 0.00000) \\
    &H&(-0.75695, 0.58588, 0.00000) \\
\end{tabular}
\end{ruledtabular}
\end{table}

\end{document}